\documentclass[finalold]{agujournal2019}
\usepackage{url} 
\usepackage{todonotes}
\usepackage[inline]{trackchanges} 
\usepackage{soul}
\usepackage{amsmath}
\usepackage{amssymb}
\usepackage{amsthm}
\usepackage{mathrsfs}

\usepackage{xcolor}
\definecolor{darkblue}{rgb}{0,0,.5}
\usepackage[colorlinks=true,
            linkcolor=darkblue, urlcolor=darkblue, citecolor=darkblue,
            raiselinks=true,
            bookmarks=true,
            bookmarksopenlevel=1,
            bookmarksopen=true,
            bookmarksnumbered=true,
            hyperindex=true,
            plainpages=false,
            pdfpagelabels=true,
            pdfstartview=FitH,
            pdfstartpage=1,
            pdfpagelayout=OneColumn]{hyperref}
\usepackage{mathtools}
\DeclarePairedDelimiter{\diagfences}{(}{)}
\newcommand{\diag}{\operatorname{diag}\diagfences}

\usepackage{multirow}

\usepackage{natbib}

\usepackage{threeparttable}

\usepackage{array}
\newcolumntype{P}[1]{>{\arraybackslash}p{#1}}
\newcolumntype{M}[1]{>{\centering\arraybackslash}m{#1}}

\usepackage[inline]{trackchanges}
\addeditor{ZN}

\usepackage{verbatim}

\draftfalse

\journalname{JGR: Solid Earth}

\begin{document}

%
%
\title{Modeling and Quantifying Parameter Uncertainty of Co-seismic Non-classical Nonlinearity in Rocks}


%
%




\authors{Zihua Niu\affil{1}, Alice-Agnes Gabriel\affil{2,1}, Linus Seelinger\affil{3}, Heiner Igel\affil{1}}

 \affiliation{1}{Department of Earth and Environmental Sciences, Ludwig-Maximilians-Universit\"at M\"unchen, Munich, Germany}
\affiliation{2}{Scripps Institution of Oceanography, UC San Diego, La Jolla, USA}
 \affiliation{3}{
Institute for Applied Mathematics, Heidelberg University, Heidelberg, Germany}




\correspondingauthor{Zihua Niu}{zniu@geophysik.uni-muenchen.de}






\begin{keypoints}
\item We analyze two physical models suitable for simulations of nonlinear elastic wave propagation observed in the laboratory.
\item The experimentally observed co-seismic acoustic modulus drop can be explained with nonlinear damage models.
\item We use the Markov chain Monte Carlo method to explore connections and uncertainties of nonlinear parameters.
\end{keypoints}

%
%

%
%


\begin{abstract}
Dynamic perturbations reveal unconventional nonlinear behavior in rocks, as evidenced by field and laboratory studies.
During the passage of seismic waves, rocks exhibit a decrease in elastic moduli, slowly recovering after.
Yet, comprehensive physical models describing these moduli alterations remain sparse and insufficiently validated against observations.
Here, we demonstrate the applicability of two physical damage models - the internal variable model (IVM) and the continuum damage model (CDM) - to provide quantitative descriptions of nonlinear co-seismic elastic wave propagation observations.
We recast the IVM and CDM models as nonlinear hyperbolic partial differential equations and implement 1D and 2D numerical simulations using an arbitrary high-order discontinuous Galerkin method.
We verify the modeling results with co-propagating acousto-elastic experimental measurements.
We find that the IVM time series of P-wave speed changes correlate slightly better with observations, while the CDM better explains the peak damage delay relative to peak strain.
Subsequently, we infer the parameters for these nonlinear models from laboratory experiments using probabilistic Bayesian inversion and 2D simulations.
By adopting the Adaptive Metropolis Markov Chain Monte Carlo method, we quantify the uncertainties of inferred parameters for both physical models, investigating their interplay in  70,000 simulations.
We find that the damage variables can trade off with the stress-strain nonlinearity in discernible ways.
We discuss physical interpretations of both damage models and that our CDM quantitatively captures an observed damage increase with perturbation frequency. 
Our results contribute to a more holistic understanding of non-classical non-linear damage with implications for co-seismic damage and post-seismic recovery after earthquakes.

\end{abstract}

\section*{Plain Language Summary}

Rocks react to earthquakes by softening when seismic waves - the energy released by earthquakes - pass through them. Observations of such rock softening during the passage of seismic waves are common both in the laboratory and in the field. Interestingly, rocks gradually harden again once the shaking stops. Different physical mechanisms have been proposed to explain the observations. 
In this study, we put two existing theories to the test, implementing them into a powerful simulation program called ExaHyPE. This allows us to model how waves move through rocks. When we compare the computer simulation outcomes with real laboratory tests, we find that both models accurately match what we see in reality.
Studying thousands of simulations with different model parameters, we find some intriguing insights. For instance, the initial state of strain and the tiny cracks that open and close within the rock may be key to understanding the hardening and softening process.
We hope to use these computer models in future earthquake simulations, offering more accurate predictions of how our Earth's crust reacts to earthquakes.

\section{Introduction}








Materials with micro- and mesoscale 
heterogeneities, such as rocks and cementitious materials,  have been widely observed to exhibit non-classical nonlinearity \citep{guyer1999nonlinear,van2000nonlinear,johnson2005slow}. Non-classical nonlinearity refers to the behavior of rocks that cannot be explained with established nonlinear stress-strain relationships, e.g. Landau's law \citep{landau1986theory}. One type of such non-classical nonlinearity can be characterized by hysteresis models \citep{preisach1935magnetische,mayergoyz1985hysteresis,mccall1996new}. Hysteresis models describe stress as a function of not only the current strain state, but of the strain history as well. Aside from hysteresis, researchers noticed that, under continuous dynamic external loads, the mechanical responses of rocks may change over time. Experimental observations of such non-classical nonlinearity were first reported in \citet{ten1996slow} from Nonlinear Resonance Ultrasound Spectroscopy (NRUS) experiments. In their NRUS, the sample's dynamic response to the excitation at the same frequency is unexpectedly different depending on whether it is measured during the upward or downward sweep across different frequencies. 
We refer to \citet{tencate2011slow} and references therein for a comprehensive summary of the NRUS observations of the latter type of non-classical nonlinearity.

\begin{figure}[htpb]
    \centering
    \includegraphics[width=0.9\columnwidth]{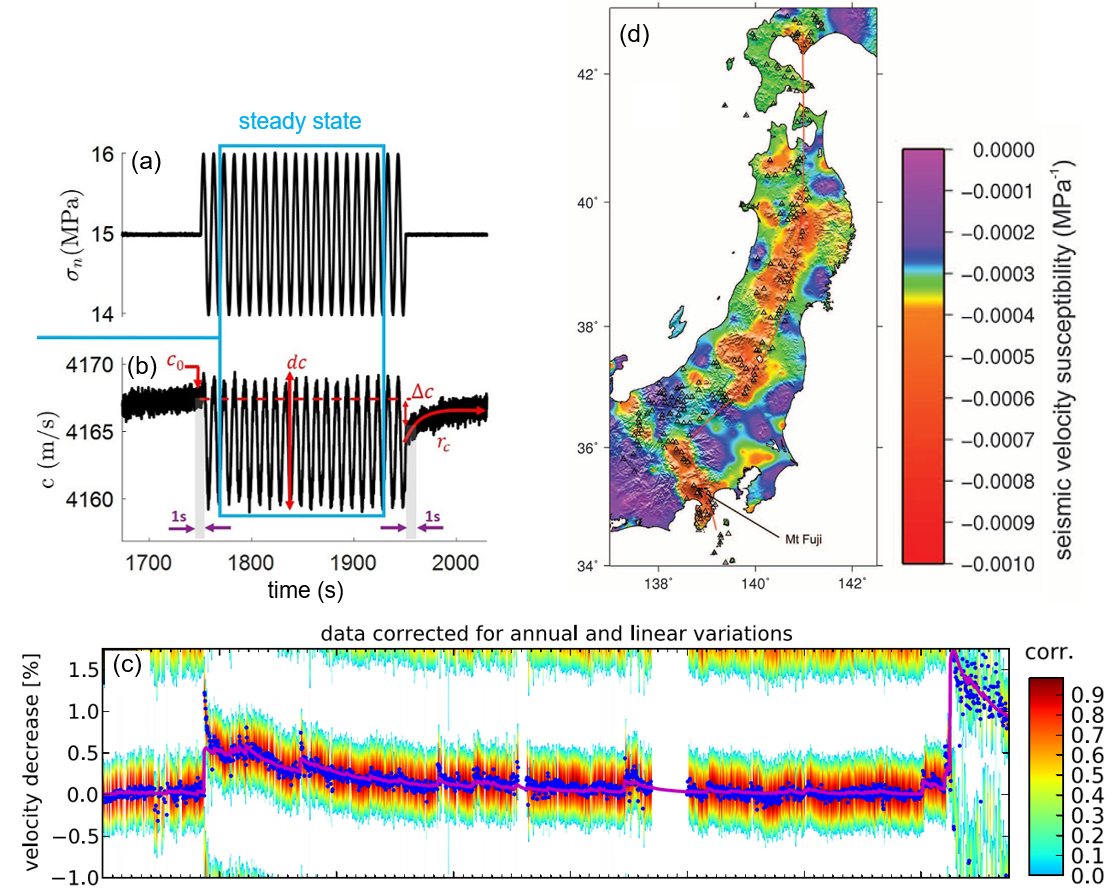}
    \caption{\small Laboratory and field observations of wave-induced non-classical nonlinearity. (a-b) Dynamic acousto-elastic testing (DAET) by \citet{manogharan2022experimental}. The sample is loaded with single-frequency dynamic stress ($\sigma_n$) perturbations shown in (a). The evolution of the change in the P-wave velocity $c$ during and after the excitation is shown in (b). (c): Monitoring of the changes in seismic velocity around station PATCX \citep{IPOC2006} in Chile between 2007 and 2014 \citep{gassenmeier2016field}. The red arrow marks the 2007 $M_w$ 7.7 Tocopilla earthquake, and the black arrows mark the occurrences of earthquakes with magnitudes between $M_w$ 5 and $M_w$ 7. (d): Map of seismic velocity drop after the 2011  $M_w$ 9.1 Tohoku-Oki earthquake \citep{brenguier2014mapping}. }
    \label{obs-slowDynamics}
\end{figure}

Details on how rock-moduli change during dynamic perturbations are first reported by \citet{renaud2012revealing} who performed Dynamic Acousto-elastic Testing (DAET) experiments on Barea sandstone. Since then, DAET experiments have been developed and conducted on different types of rock samples  \citep[e.g.,][]{riviere2013pump,riviere2015set,jin2018dynamic}. Figs. \ref{obs-slowDynamics}a and b show one of the  most recent DAET measurements by \citet{manogharan2022experimental} on Westerly granite rock samples under triaxial loading conditions.\citet{manogharan2022experimental} load the sample with 20 cycles of 0.1 Hz sinusoidal stress perturbations in one direction. They monitor the change of P-wave speeds during the oscillations and also after the oscillation stops.
In response to the imposed stress oscillations, the wave speed evolves in three consecutive phases (Fig. \ref{obs-slowDynamics}b): (1) As loading starts, the wave velocity of the sample reaches a non-equilibrium state and experiences an overall drop aside from fluctuations with the cyclic loading. This phase is referred to as the conditioning of the material; (2) The overall drop of wave velocity stabilizes and reaches a new steady state; (3) After removing the cyclic loading, the velocity recovers over an extended period of time. We will refer to all three phases as ``slow dynamics'' in the following.


Similar observations have also been reported from the field. \citet{gassenmeier2016field} monitor the change of seismic velocity around the station PATCX from the Integrated Plate Boundary Observatory Chile Network (IPOC) \cite{IPOC2006}. During the occurrence of the $M_w$ 7.7 Tocopilla earthquake in Chile on November 14th, 2007 (marked with the red arrow in Fig. \ref{obs-slowDynamics}c), \citet{gassenmeier2016field} observe a drop and subsequent recovery of seismic velocity within a radius of $\approx$ 2.3~km surrounding a station that is about 100 km away from the fault using coda wave interferometry in the frequency range of 4-6~Hz. Similar co-seismic velocity drops are reported to occur not only during large earthquakes but also due to intermediate earthquakes with magnitudes between $M_w$ 5 and $M_w$ 7 (marked with black arrows in Fig. \ref{obs-slowDynamics}c). \citet{brenguier2014mapping} map the co-seismic velocity drop in Japan after the 2011 Tohoku-Oki earthquake (see Fig. \ref{obs-slowDynamics}d) using data from the high sensitivity seismograph network (Hi-net) in Japan \citep{takanami2003hi}. While it is commonly assumed that transient fluid effects are key to the observed velocity drop \citep{brenguier2014mapping,illien2022seismic}, \citet{manogharan2021nonlinear,manogharan2022experimental} find that a velocity drop during dynamic perturbations of fractured fluid-saturated rocks is not more significant than that of fractured dry rocks. 

Despite phenomenologically comparable observations, it is currently unclear how relevant the non-classical nonlinearity observed in laboratory experiments is for observed co-seismic velocity drops in the field and if an overarching theoretical physical framework can be established. Furthermore, despite the relatively low velocity drops (usually smaller than 1\%) using ambient noises and coda wave Interferometry \citep[e.g.,][]{brenguier2014mapping,gassenmeier2016field,illien2022seismic}, more recent measurements based on auto-correlation of a single station \citep{lu2022regional} or the combination of surface and borehole stations \citep{qin2020imaging,wang2021near}, observed velocity drop can be higher than 10\%. Unifying laboratory and field observations is challenged by the vastly different scales and sparsity of high-resolution observations. Numerical models can help overcome these challenges. 
As a first step, we here focus on identifying appropriate theoretical model(s) that can be informed and verified by recent laboratory experiments and are applicable for future numerical models on the field-scale of observations. 

In (computational) seismology, a range of models has been proposed to explain or predict source and site effects of nonlinearity on earthquake nucleation, rupture dynamics, and ground motions. The Masing-Prager-Ishlinski-Iwan (MPII) model \citep{iwan1967class} has been implemented to study local non-linear site effects due to soft sediments \citep{roten2013high,roten2018implementation,oral20192,oral2022kathmandu}. The MPII model reproduces the hysteretic stress-strain relations and can be extended to explain the excess-pore pressure in liquefiable soils \citep{oral20192}. The nonlinearity of fault zone deformation has been modeled as co-seismic off-fault brittle continuum damage \citep[e.g.,][]{xu2015dynamic,ThomasBhat2018} or non-associative Drucker-Prager off-fault plasticity \citep[e.g.,][]{andrews2005,wollherr2018off}, explicit secondary tensile and shear fracturing \citep{okubo2020modeling, yamashita2000, gabriel2021unified}, using a volumetric representation of fault zones governed by reformulated rate-and-state friction laws \citep{preuss2019seismic,preuss2020characteristics,pranger2022rate} or phase-field inspired methods \citep[e.g.,][]{fei2023phase}. In this work, we do not focus on brittle damage since we are interested in path nonlinearity which is associated with stress levels well below yield strength.

The first model that attempted to explain non-classical nonlinearity focused on the possible physical processes that might exist at the microscopic scale.
\citet{delsanto2003modeling} proposed a model that includes thermally activated random transitions between two different interstitial states. Since then, many studies use the framework of adhesive contacts at rough crack surfaces to explain slow dynamics \citep{pecorari2004adhesion,aleshin2007microcontact,lebedev2014unified,wang2021near}. The adhesion potential at the contact of the grain boundary can have two local stationary points. \citet{lebedev2014unified} argue that smaller asperities can overcome the potential barrier between the two stationary points and reach a secondary meta-stable state during the perturbations. These contacts may gradually return to the initial state of equilibrium due to thermal fluctuations and may be responsible for the slow dynamics. 

Alternatively, the apparent logarithmic recovery of material moduli with time may be explained by a superposition of exponential evolution processes at different time scales. This idea is supported both mathematically \citep{snieder2017time,sens2019model} and from experimental observations \citep{shokouhi2017slow}. The anisotropic elastic moduli drop can then be to first order explained by a scalar conditioning variable \citep{lott2017nonlinear} and the corresponding stiffness tensor has been defined by \citet{hughes1953second}. 

While the above physical models are firmly rooted in processes at the microscopic scale, they introduce a significant number of parameters. Many of these parameters are hard to constrain directly from observations. Also, the proposed models are restricted to 0D (oscillation) or 1D analysis, and extension to 2D and 3D  would require a substantial effort although it is crucial for the verification and interpretation of field observations. There also exist phenomenological models, with fewer parameters, that have been developed over the last two decades, such as the soft ratchet model of \citet{vakhnenko2004strain,vakhnenko2005soft}. Their model includes a fast subsystem of displacement and a slow subsystem of ruptured intergrain cohesive bonds. The concentration of the ruptured bonds is represented by an internal (damage) variable that evolves to a stress-dependent equilibrium. \citet{FAVRIE2015221} fully couple these two subsystems and proposed a numerical method to solve the coupled equations based on the Finite Volume method. 

\citet{berjamin2017nonlinear} follows the phenomenological soft ratchet approach by using one internal variable to describe the evolution of elastic material moduli. This model will be herein referred to as the internal variable model (IVM) in the following. IVM ensures that the kinematics of the internal variable complies with the laws of thermodynamics. The model can reproduce the conditioning and recovery phases of nonclassical nonlinear elasticity with only two additional parameters than used in classical nonlinearity. However, their expression for the internal energy of the material contains a term that is not clearly linked to any physical process. 

In this work, we offer a physics-based understanding of the origin of this phenomenological term.
%
To this end, we resort to continuum damage mechanics (CDM) \citep{kachanov1958creep,kachanov1986introduction}. We show that the internal variable in \citet{berjamin2017nonlinear} is very similar to a scalar damage variable in typical CDM approaches \citep{chaboche1988continuum}. The scalar damage variable approach is popular in diffuse interface approaches, e.g., to ‘smear out’ sharp discontinuous cracks via a smooth but rapid transition between intact and fully damaged material states \citep{borst2004,tavelli2020}. In the CDM context, the damage variable is a mathematical representation of defects, e.g. intergranular cavities or microcracks density, which are distributed in a solid. Different CDM models are characterized by different assumptions on the distribution of microscopic defects and by the various mathematical operations of homogenization. \citet{budiansky1976elastic} derive the elastic moduli as a function of the damage variable with the homogenization of cracks which directions are uniformly and omnidirectionally distributed. The Godunov–Peshkov–Romenski (GPR) model \citep{resnyansky2003constitutive,romenski2007conservative,tavelli2020,gabriel2021unified} uses a different homogenization scheme, continuum mixture theory \citep{romenski2004compressible}, to define a material that is a mixture of a "totally-damaged" and an "undamaged" constituents. 

However, including the above two models of \citet{budiansky1976elastic} and \citet{gabriel2021unified}, the internal energy in most CDM-based damage models drops with the increase of the damage variable. This prevents the possibility of healing. The continuum damage healing model (CDHM) introduces an additional healing variable aside from the damage variable \citep{darabi2012continuum,oucif2018continuum}. A potentially simpler framework that can allow healing is the damage model proposed by \citet{lyakhovsky1997distributed}. It is based on homogenizing micro-cracks oriented perpendicular to the maximum tension (or compression). The internal energy of this model is not unconditionally decreasing with the increase of the damage variable. However, it has not yet been investigated how this model can be related to observed slow dynamics.

We summarize selected representative damage models with respect to non-classical nonlinearity in Table \ref{modelscomparison}. We compare these models in terms of four characteristics: (1) Stress-strain relation under zero damage; (2) Whether the model differentiates between extension and compression in damage kinematics; (3) Linear or nonlinear relationship between the damage variable and the moduli; (4) Whether they contain the mechanism of healing.

\begin{table}[ht]
\centering
\caption{\small Summary of chosen representative damage models (see text for details). The two models that we focus on in this work are highlighted in bold.}
\begin{threeparttable}
\begin{tabular}{ P{1cm} P{2.5cm} P{2.5cm} P{2.5cm} P{2.5cm}}
\hline
models &stress-strain relationship before damage &damage kinematics in compression and extension &relationship between damage variable and moduli & healing mechanisms\\
\hline
\textbf{IVM\tnote{1}} &Nonlinear &Same &Linear &Yes\\
\textbf{CDM\tnote{2}} &Linear &Different &Linear &Yes\\
CDM\tnote{3} &Linear &Same &Nonlinear &No\\
GPR\tnote{4} &Linear &Same &Nonlinear &No\\
CDHM\tnote{5} &Linear &Same &Nonlinear &Yes\\
\hline
\end{tabular}
\begin{tablenotes}\footnotesize
\item[1] Model B, \citet{berjamin2017nonlinear}
\item[2] Model L, \citet{lyakhovsky1997distributed}
\item[3] \citet{budiansky1976elastic}
\item[4] \citet{resnyansky2003constitutive,romenski2007conservative,tavelli2020,gabriel2021unified}
\item[5] \citet{darabi2012continuum}
\end{tablenotes}
\end{threeparttable}
\label{modelscomparison}
\end{table}

In the following, we will focus on two models, IVM by \citet{berjamin2017nonlinear} and CDM by \citet{lyakhovsky1997distributed}, and will refer to them as ``Model B'' and ``Model L'', respectively, for brevity. We demonstrate their unique capabilities to describe the observed wave speed changes in laboratory experiments.
We compare both models and show that the CDM nonlinear damage model proposed by \citet{lyakhovsky1997distributed} can also explain slow dynamics. We implement both models in the arbitrary high-order discontinuous Galerkin (ADER-DG) solver ExaHyPE \citep{reinarz2020exahype} and verify the numerical simulation results with experimental observations. We infer model parameters from laboratory measurements using a Markov chain Monte Carlo (MCMC, \cite{MCMC}) algorithm. This takes uncertainties due to measurement errors into account and allows us to investigate model parameters' relative importance and their interactions. Lastly, we discuss that the CDM model by \citet{lyakhovsky1997distributed} may be a preferred model for large-scale wave propagation simulations capable of linking observations of co-seismic damage in the field with laboratory findings and continuum damage mechanics theory.
%
%

In Section \ref{Method}, the two models of \citet{lyakhovsky1997distributed} and \citet{berjamin2017nonlinear} (Model L and Model B hereafter) are summarized within the framework of thermodynamics. We propose a way to explain the origin of the phenomenological term in the model of \citet{berjamin2017nonlinear}. We then describe the numerical simulation of the nonlinear wave propagation with the two models. This is followed by a description of the experiment that we will compare to the Bayesian problem used for parameter inference and the Adaptive Metropolis Markov Chain Monte Carlo (AM-MCMC) method solving it.  In Section \ref{Results}, we compare the two models regarding the damage evolution during dynamic acousto-elastic testing and measured amplitude- and frequency-dependent damage. The inversion results of the model parameters are also shown. Finally, the performance and restrictions of the two models will be discussed in Section \ref{Conclusions and outlook}. 








\section{Method}
\label{Method}

\subsection{The thermodynamic formulation of nonlinear damage models}

In the framework of continuum damage mechanics, a scalar variable can describe the changes of elastic moduli with damage. 
We start with the 1st law of thermodynamics,

\begin{equation}
   \overset{\cdot}{e} = \overset{\cdot}{w} + \overset{\cdot}{q},
   \label{thermo1st}
\end{equation}
where $\overset{\cdot}{(\cdot)}$ denotes the time derivative, $e$ is the specific internal energy of the system normalized by mass, $w$ is the external work  per unit volume of the system and $q$ is the absorbed heat from the environment  per unit volume of the system. At the time scale of elastodynamic processes, the heat transfer and any possible heat sources are assumed to be negligible, i.e. we assume an adiabatic process where $\overset{\cdot}{q} = 0$. In case of only considering mechanical work, it is $\overset{\cdot}{w} = \underset{=}{\sigma} : \overset{\cdot}{\underset{=}{\epsilon}}$ and $\underset{=}{(\cdot)}$ denotes a tensor of rank two.

The expression of the internal energy depends on the choice of state variables that are used to describe the system. For an elastic material, we chose the strain $\underset{=}{\varepsilon}$ and the specific entropy $s$ as state variables. In addition, to incorporate the damage to the material, another scalar state variable $\alpha$ is included. This means $e \equiv e(s, \underset{=}{\varepsilon},\alpha)$.

With the above definitions of state variables, the Gibbs identity can be written as

\begin{equation}
    \overset{\cdot}{e} = T \overset{\cdot}{s} + \dfrac{\partial e}{\partial \underset{=}{\varepsilon} } \overset{\cdot}{\underset{=}{\epsilon}} + \dfrac{\partial e}{\partial \alpha } \overset{\cdot}{\alpha},
    \label{gibbs}
\end{equation}
where $T=\dfrac{\partial e}{\partial s}>0$ is the absolute temperature. 

Different nonlinear or damage models have different ways of defining the internal energy as a function of  $\underset{=}{\varepsilon}$ and $\alpha$.
The combination of Eqs. (\ref{thermo1st}) and (\ref{gibbs}), together with the earlier defined assumptions of $\overset{\cdot}{q} = 0$ and $\overset{\cdot}{w} = \underset{=}{\sigma} : \overset{\cdot}{\underset{=}{\varepsilon}}$, yields

\begin{equation}
    T\overset{\cdot}{s} = (\underset{=}{\sigma} - \dfrac{\partial e}{\partial \underset{=}{\varepsilon} }) \overset{\cdot}{\underset{=}{\epsilon}} - \dfrac{\partial e}{\partial \alpha } \overset{\cdot}{\alpha}.
    \label{non-eq}
\end{equation}

For a spontaneous process in an adiabatic system, $\text{d}s \geq 0$ for any given $\overset{\cdot}{\varepsilon}$ and $\overset{\cdot}{\alpha}$, which is known as the Clausius–Duhem inequality \citep{truesdell1952mechanical}. Assuming $\underset{=}{\sigma}$ is independent of $\overset{\cdot}{\varepsilon}$, we derive that

\begin{equation}
    \underset{=}{\sigma} = \dfrac{\partial e}{\partial \underset{=}{\varepsilon} },
    \label{stress-strain}
\end{equation}
and
\begin{equation}
    \dfrac{\partial e}{\partial \alpha } \overset{\cdot}{\alpha} \leq 0,
    \label{disspate-alpha}
\end{equation}
where $\dfrac{\partial e}{\partial \alpha }$ can be any function of the state variables. Eq. (\ref{disspate-alpha}) describes the evolution of the damage variable $\alpha$. A simple and non-trivial (non-zero) expression can be

\begin{equation}
    \dfrac{\partial e}{\partial \alpha } = - \tau \overset{\cdot}{\alpha},
    \label{damage-evolu}
\end{equation}
where $\tau$ can be any non-negative constant or non-negative function of the state variables. Substituting Eq. (\ref{damage-evolu}) and (\ref{stress-strain}) into Eq. (\ref{non-eq}) yields the energy dissipation rate of the system,

\begin{equation}
  \mathscr{D} = T \overset{\cdot}{s} =
    \begin{cases}
      0 
      & \text{if $\tau$ = 0}\\
      \frac{1}{\tau} (\dfrac{\partial e}{\partial \alpha })^2 
      & \text{if $\tau >$ 0}\\
    \end{cases}.
    \label{energy dissipation}
\end{equation}

Here damaged material will only heal when $\tau > 0$ and $\dfrac{\partial e}{\partial \alpha} < 0$; otherwise damage will steadily accumulate. Both Model L \citep{lyakhovsky1997distributed} and Model B \citep{berjamin2017nonlinear} describe the mechanisms of damage and recovery. But they are introduced under different assumptions regarding the form of the internal energy as a function of strain $\underset{=}{\varepsilon} $ and the damage variable $\alpha$.

The internal variable of \citet{berjamin2017nonlinear} is defined as

\begin{equation}
    e = (1-\alpha) W(\underset{=}{\varepsilon}) + \phi(\alpha),
    \label{inter-energy-berjamin}
\end{equation}
where $\phi(\alpha)$ is called the storage energy and increases with the development of damage. 
In the 1D case where only $\varepsilon_{xx} = \varepsilon$ is non-zero, \citet{berjamin2017nonlinear} express $W$ as $W = (\dfrac{1}{2} - \dfrac{\beta}{3} \varepsilon - \dfrac{\delta}{4} \varepsilon^2) E \varepsilon^2$ based on Landau's law \citep{landau1986theory}. 
In the 2D plane-strain case, \citet{berjamin2019plane} use the Murnaghan's law $W = \dfrac{\lambda+2 \mu}{2}E_I^2 - 2 \mu E_{II} + \dfrac{l+2m}{3} E_I^3 - 2 m E_I E_{II} + n E_{III}$ \citep{murnaghan1937finite}. $l$, $m$, and $n$ are the three Murnaghan coefficients (third-order elastic constants), while $E_I$, $E_{II}$ and $E_{III}$ are three stress invariants that are defined in \citet{berjamin2019plane}.

According to Eq. (\ref{damage-evolu}), the damage evolution then reads

\begin{equation}
    \overset{\cdot}{\alpha} = \dfrac{1}{\tau}
    (W(\underset{=}{\varepsilon}) - \phi'(\alpha)),
    \label{damage-evolu-berjamin}
\end{equation}
where \citet{berjamin2017nonlinear} proposed two possible expressions of $\phi$: $\phi'(\alpha) = \gamma_b \dfrac{\alpha}{1-\alpha^2}$ or $\phi'(\alpha) = \gamma_b \alpha$. The scaling factor $\tau$ is formulated as $\tau = \gamma_b \tau_b$, where $\gamma_b$ is the scale of storage energy $\phi(\alpha)$ and $\tau_b$ is the time scale of damage evolution.

In the Model L proposed by \citet{lyakhovsky1997distributed} internal energy is defined as

\begin{equation}
    e = W(\underset{=}{\varepsilon},\alpha) - \gamma I_1 \sqrt{I_2}
    \label{inter-energy-lya},
\end{equation}
where, in the linear elastic case, $W(\underset{=}{\varepsilon},\alpha) = \dfrac{\lambda}{2} I_1 + \mu I_2$, $I_1 = \varepsilon_{kk}$ and $I_1 = \varepsilon_{ij} \varepsilon_{ij}$ are the first and the second strain invariant, and $\gamma$ is a third modulus that originates from the homogenization of parallel cracks \citep{lyakhovsky1997non}. It is assumed that $\lambda = \lambda_0$, $\mu = \mu_0 - \alpha \mu_r$ and $\gamma = \alpha \gamma_r$. The corresponding damage kinematics then read

\begin{align}
    \overset{\cdot}{\alpha} &= C_d \gamma_r
    (\xi_0 I_2 + I_1 \sqrt{I_2}) \nonumber \\
    &= C_d \gamma_r I_2 
    (\xi_0  + \xi),
    \label{damage-evolu-lya}
\end{align}
where $\xi_0 = \dfrac{\mu_r}{\lambda_r}$, $\xi = I_1/\sqrt{I_2}$ and $C_d$ can be any non-negative function of the state variables $\underset{=}{\epsilon}$ and $\alpha$. We note that the material heals when $\xi_0  + \xi < 0$ and is damaged when $\xi_0  + \xi > 0$. 

A comparison of Eqs. (\ref{damage-evolu-berjamin}) and (\ref{damage-evolu-lya}) shows that both models in principle include the mechanics of healing. However, the crucial term $\phi$ for explaining slow dynamics in Model B in Eq.(\ref{inter-energy-berjamin}) has only limited physical meaning, which challenges the interpretation of the physical mechanisms driving the observed slow dynamics (Fig. \ref{obs-slowDynamics}). On the other hand, conditioning and healing were not yet explored using the model L of \citet{lyakhovsky1997distributed}. 

%
So far we have summarized Model L and Model B in the same thermodynamics framework.
Next, we explain why, based on Model L, it may be physically plausible to include a term in the internal energy $W$ that can increase with damage and how this term can result in both the conditioning during dynamic perturbations and the recovery after removing the perturbations.
We here propose that the stationary damage during perturbations can be recovered by assuming the following form of damage kinematics in Eq.(\ref{damage-evolu-lya-ada}) for the Model L that differentiates the evolution laws during the recovery from that during the damage.
A recovery phase after dynamic perturbations can occur if an initial strain that satisfies $\xi_0  + \xi < 0$ exists. This strain level does not need to be large since $\xi$ is only related to the relative magnitude of each strain component (i.e., the shape of the strain tensor).

\begin{equation}
  \overset{\cdot}{\alpha} = 
    \begin{cases}
      C_d(\alpha) \gamma_r I_2 (\xi_0  + \xi) 
      & \text{if $\xi_0  + \xi >$ 0 and $\alpha \geq$ 0}\\
      C_r(\alpha) \gamma_r I_2 (\xi_0  + \xi)
      & \text{if $\xi_0  + \xi \leq$ 0 and $\alpha \geq$ 0}\\
      0
      & \text{if $\alpha <$ 0}\\
    \end{cases}.
    \label{damage-evolu-lya-ada}
\end{equation}
The evolution laws during `damage' ($C_d(\alpha)$, the conditioning during dynamic perturbations) and recovery ($C_r(\alpha)$) are treated separately. Stationary damage requires the increase of $C_d(\alpha)$, the decrease of $C_r(\alpha)$ or both. At the same time, $C_d(\alpha)$ and $C_r(\alpha)$  are required to remain non-negative to preserve a non-decreasing entropy of the system according to Eq. (\ref{energy dissipation}). With the above considerations, we choose $C_d(\alpha) = c_d \exp{(-\dfrac{\alpha}{\alpha_d})}$ and $C_r(\alpha) = c_r \alpha$ to represent, respectively, the decrease of damage coefficient $C_d$ and the increase of recovery coefficient $C_r$ with the increase of the damage variable. 


Our proposed damage evolution still follows the laws of thermodynamics. We show in Section \ref{Results} that the combination of Eq. (\ref{damage-evolu-lya-ada}) and the existence of a well-defined initial strain level recovers many aspects of the observed slow dynamics. 

\subsection{Numerical implementation}
\begin{figure}[htpb]
    \centering
    \includegraphics[width=0.9\columnwidth]{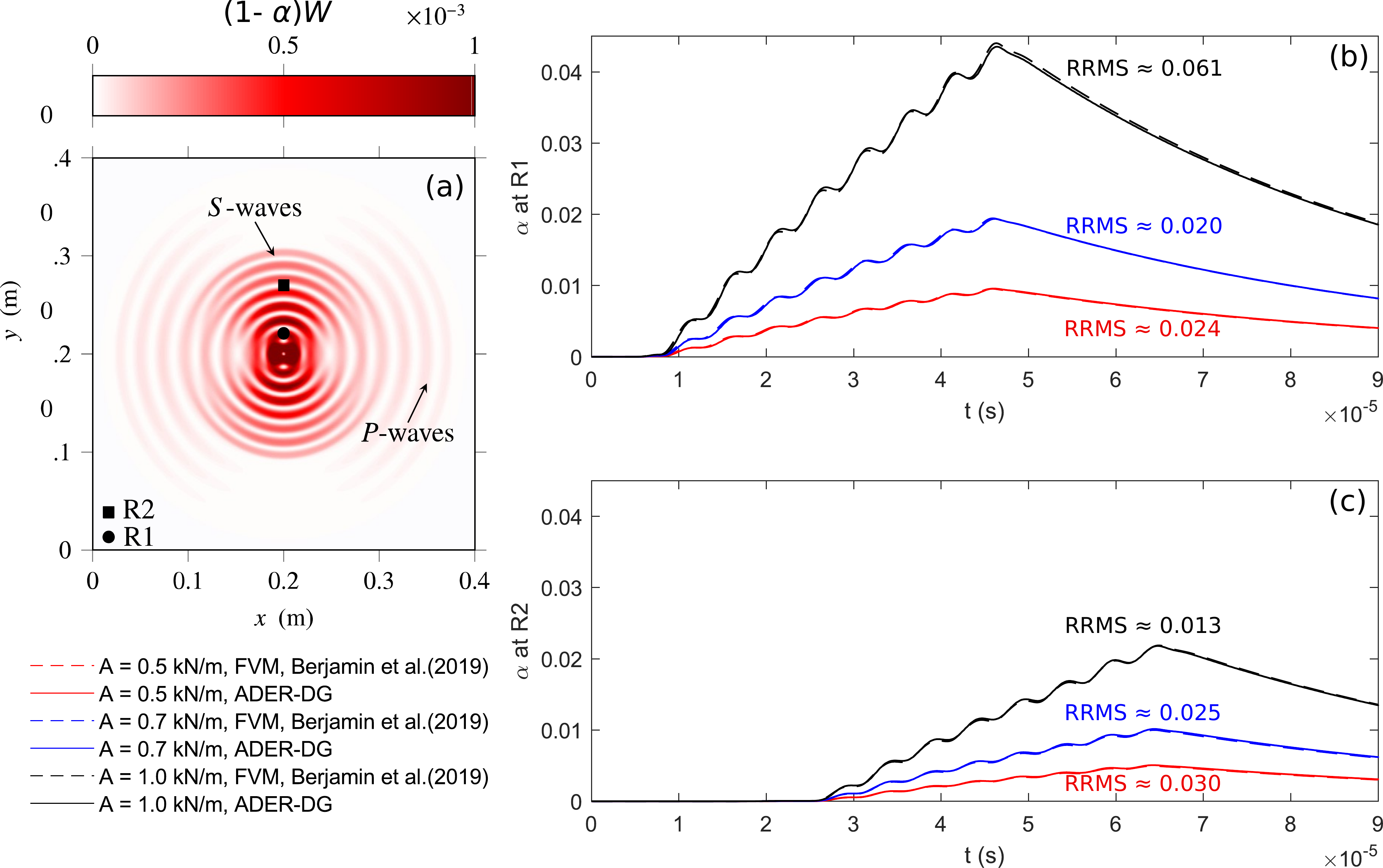}
    \caption{\small (a) Snapshot of the elastic strain energy field $(1-\alpha) W \text{in J}/\text{m}^3 $ at $t=0.04$ ms by \citet{berjamin2019plane} simulated using the Finite Volume Method (FVM). R1 at (0.2, 0.22) m and R2 at (0.2, 0.27) m are the locations of two receivers where the recorded time series of the damage variable $\alpha$ is shown in (b) and (c), respectively. The solutions from \citet{berjamin2019plane} with FVM are plotted as dashed curves and the ExaHyPE solutions of our implementation with the arbitrary high-order discontinuous Galerkin (ADER-DG) method are plotted in solid curves for R1 in (b) and R2 in (c). Different source amplitudes with $A$ = 0.5, 0.7 and 1.0 kN/m are plotted in red, blue and black, respectively. The relative root mean square (RRMS) errors = 
    $\sqrt{(\boldsymbol{\alpha_1}-\boldsymbol{\alpha_2})^2/(\boldsymbol{\alpha_2}-\overset{-}{\boldsymbol{\alpha_2}})^2}$, derived from the solution vectors of ADER-DG $\boldsymbol{\alpha_1}$ and FVM $\boldsymbol{\alpha_2}$ respectively are denoted and $\overset{-}{\boldsymbol{\alpha_2}}$ is the average of $\boldsymbol{\alpha_2}$. 
    }
    \label{DG2D}
\end{figure}

We implement both models in ExaHyPE \citep{reinarz2020exahype}, an engine built for solving nonlinear hyperbolic partial differential equations (PDEs) with the arbitrary high-order derivative discontinuous Galerkin (ADER-DG) method.
%
%
We verify the implementation by means of comparison of our numerical solutions with those given by \citet{berjamin2019plane}, who implemented the IVM B model in 2D under plain-strain conditions using the finite volume method with flux limiters. 
In this verification benchmark, the simulation domain is [0.0, 0.4]~m $\times$ [0.0, 0.4]~m. A point force radiates seismic waves and is defined as $f_x=A \sin{2 \pi f_c t} \delta(x-x_s) \delta(y-y_s)$, where $\delta$ is the Dirac delta function injected at (0.2, 0.2)~m in x-direction and $f_c$ = 100~Hz. Following the implementation of \citet{berjamin2019plane}, the Dirac delta function is approximated as

\begin{equation}
    \delta(x-x_s) \delta(y-y_s) = \dfrac{ \exp{(-(d/\sigma_c)^2)} }{ \pi \sigma_c^2 (1-\exp{ (-(R/\sigma_c)^2) }) } \textbf{1}_{d \leq R},
    \label{dirac-app}
\end{equation}
where $d = \sqrt{ (x-x_s)^2 + (y-y_s)^2 }$; $\textbf{1}_{d \leq R}$ is the indicator function whose support is a disk space with a radius of $R = c_p / (7.5 f_c)$; $c_p = \sqrt{(\lambda + 2 \mu)/rho_0}$ is the speed of the P-wave in the undamaged material and the width parameter of the Gaussian function is $\sigma_c = R/2$. In the simulation using the ADER-DG method in ExaHyPE, a structured quadrilateral computational mesh is used to discretise space with an element edge length of around 1.66~mm, which means that 27 cells resolve one wavelength of P waves and 16 cells resolve one wavelength of S waves. We choose the 1st order Lagrange basis with Gauss-Legendre quadrature nodes.
The achieved excellent agreement is shown in Fig. \ref{DG2D}.  (a) shows the map of elastic strain energy at $t=0.04$~ms from \citet{berjamin2019plane}. 
Solutions to the damage ($\alpha$) evolution at the two receivers, R1 at (0.2, 0.22) m and R2 at (0.2, 0.27) m are compared between the Finite Volume solutions of \citet{berjamin2019plane} and our implementation in ExaHyPE in Figs. \ref{DG2D}b and c. 

\subsection{Verification and parameter constraints from laboratory observations}

One of the main advantages of Model B (IVM) and Model L (CDM) is they have fewer parameters and therefore maybe easier to constrain than models that are established on detailed physical processes at microscopic scales. We now generate ensembles of our numerical simulations using the two models and compare to laboratory measurements. We apply Bayesion inversion to quantify how well the model parameters can be constrained from laboratory experiments. Most laboratory experiments of slow dynamics are based on 1D setups. \citet{feng2018short} proposed an experimental setup (copropagating acousto-elastic testing, Fig. \ref{Lab setup}) that enables the observation of acoustic modulus change during the propagation of waves in a rock sample. In the following, we first describe the experimental setup. Next, we formulate a Bayesian inversion problem that we apply for quantitative characterization of the theoretical model parameters and their associated uncertainties with respect to reproducing laboratory results. The Bayesian inversion problem is solved with a Markov chain Monte Carlo (MCMC, \citet{MCMC}) type method.

\subsubsection{Laboratory verification experiment measuring slow dynamics in 2D}
\label{Measurement in 2D setup}

\begin{figure}[htpb]
    \centering
    \includegraphics[width=1\columnwidth]{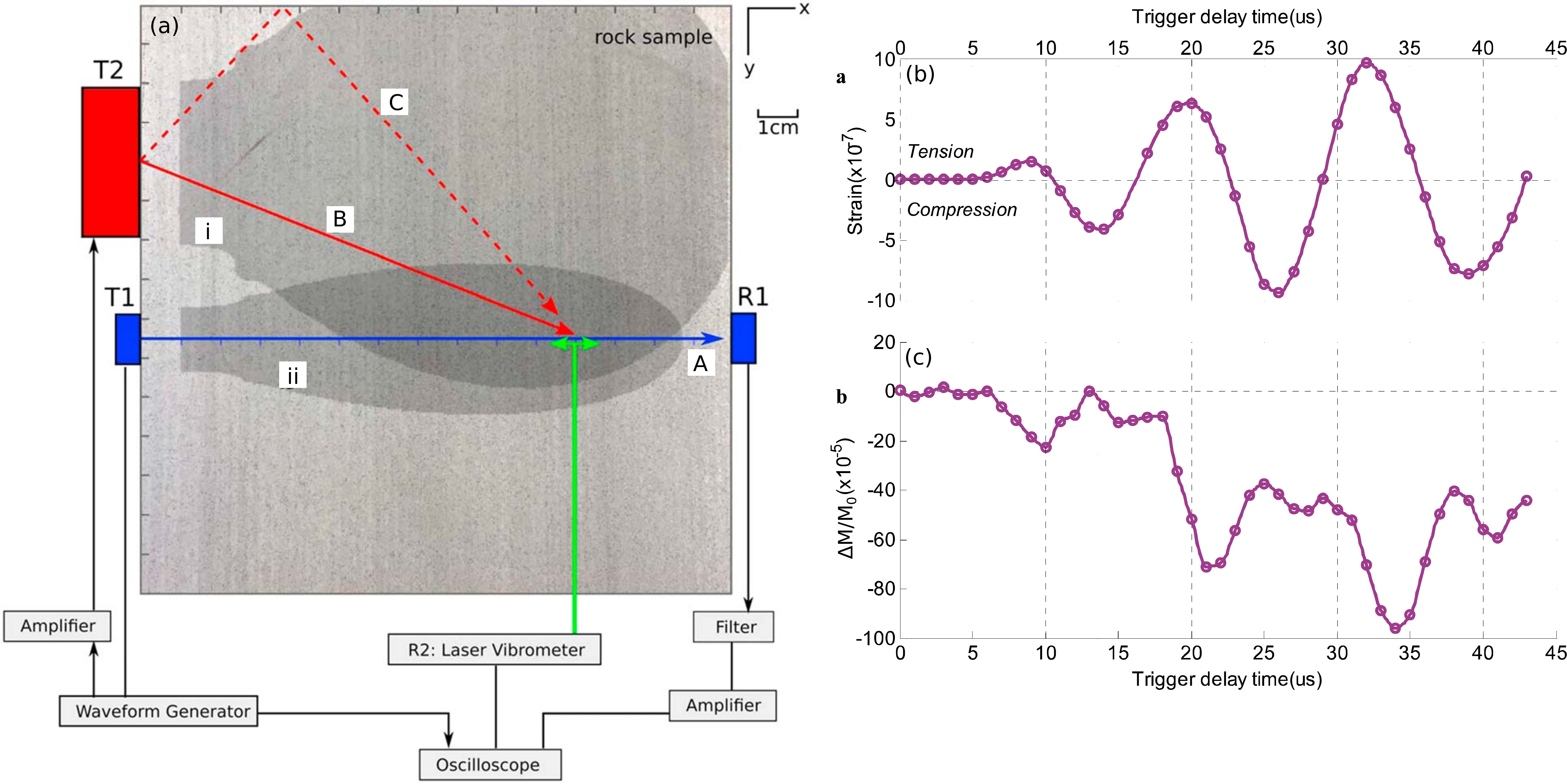}
    \caption{\small (a) Experimental setup to perform copropagating acousto-elastic testing (Figure adapted from \citet{feng2018short}). T1 is a low-energy high-frequency (HF) transmission ultrasound (US) transducer, the probe, and R1 is the receiver of T1. T2 is a high-energy low frequency (LF) transmission US transducer. The induced velocity field is recorded by the laser vibrometer R2. The velocity field is converted to the strain field with the method described by \citet{feng2018short}. The shadowed areas indicate the radiation pattern (amplitude as a function of angle) of (i) T2 and (ii) T1. Line A is the direct ray path from T1 to R1, line B is the direct ray path from T2 to a point of wave interaction at (8.5, 11.0) cm, and the dashed line C shows the ray path for the wave from T2 reflected at the top boundary of the sample that arrives at that same point of interaction at (8.5, 11.0) cm. (b) The strain ($\varepsilon_{xx}$) measured at the wavefront of the waves that are excited from T1 at different trigger time delays. Each data point is the averaged strain at the wavefront during its propagation from T1 to R1. (c) The estimated relative change in acoustic modulus, based on the travel time difference between T1 and R1, as a function of the trigger time delays. Acoustic modulus $M = \lambda + 2 \mu$. $M_0$ is the $M$ without the pumping signals, while $\Delta M$ is the change in $M$ during the propagation of the pumping signals.}
    \label{Lab setup}
\end{figure}

The experiment is conducted using a sample of Crab Orchard sandstone of size 15 cm $\times$ 15 cm $\times$ 5 cm. \citet{feng2018short} attached two ultrasound transducers and one receiver to the rock sample as shown in Fig. \ref{Lab setup}a. T1 is a low-energy high-frequency (HF) transmission ultrasound (US) transducer, the probe, and R1 is a HF reception US transducer. T2 is a high-energy low frequency (LF) transmission US transducer, the pump, and R2 is the laser vibrometer. In the experiment, T2 generates a pumping signal with a frequency of 74 kHz. The particle velocity field excited by T2 is measured with the vibrometer R2 and the particle velocity is converted to the strain ($\varepsilon_{xx}$) along the ray path A between T1 and R1. The P-wave speed along the ray path A is probed with a 620 kHz signal from T1. The amplitude of the perturbation by T1 is of much lower amplitude than that by T2 and is therefore assumed to not perturb the strain field. Once T2 is triggered at $t_0$, T1 will send signals every 1 $\mu$s to measure the P-wave speed along the ray path A. The time difference between the signal from T1 and $t_0$ is called "trigger delay". The strain field at the wavefront of each trigger delay along the ray path A is averaged and shown in Fig. \ref{Lab setup}b. The acoustic modulus change $\Delta M/M_0$ is computed from the change in P-wave speed of each trigger delay in Fig. \ref{Lab setup}c.

\citet{feng2018short} explain the relation between the strain field and the acoustic modulus change with the nonlinear visco-elastic relationship. They observe a time shift of around 2 $\mu$s between the peak in the acoustic modulus change and in strain, which is fit by imposing a ``delay time'' $\Delta t$ in the visco-elastic relationship. We show in this work that the data may also be explained with the proposed damage models, without having to impose a ``delay time'' for explaining the time shift.

\subsubsection{Probablistic inversion and uncertainty quantification with a Markov chain Monte Carlo approach}

We begin by defining configurations of our two competing deterministic models, matching the experimental setup. Next, we augment the deterministic models by embedding them in a Bayesian inversion problem. This allows us to infer model parameters from data and discover their interactions, taking into account possible ambiguities and the effect of uncertainties in experimental measurements.

\paragraph{Deterministic models} \mbox{} \\

We model the 2D slow dynamics experimental setup with both the L and the B models. The perturbation from T2 is simulated as a Dirichlet boundary condition distributed in the area where T2 is in contact with the sample. The remaining boundaries are treated as free surfaces with zero traction. As in the experiment, the strain $\varepsilon_{xx}$ and the acoustic modulus change are averaged over the path of the wavefront between T1 and R1 for each trigger delay (every 1 $\mu$s between 0 and 39 $\mu$s).

In Eq. (\ref{inter-energy-berjamin}) of model B, the first order nonlinearity in 2D Murnaghan's law already involves three parameters ($l$, $m$ and $n$). The laboratory data are not sufficient to constrain all three parameters. Similar to the simplification made in the original paper of \citet{feng2018short}, the change in $M = \lambda + 2 \mu$, which is related to the speed of the P-wave, is simplified as a function of the damage parameters computed from the 2D simulation and the 1D first and second order nonlinearity parameters as in Eq. (\ref{mod-change-1d}).

\begin{equation}
    M = M_0 (1-g) (1 - \beta \varepsilon_{xx} - \delta \varepsilon_{xx}^2),
    \label{mod-change-1d}
\end{equation}
where $M_0$ is the initial acoustic modulus of the undamaged rock sample before the perturbations. Such a relationship is also used by \citet{berjamin2017nonlinear} in the 1D formulation of their model. For model L, we derive the variations in speed of the P-wave with the strain tensor and the damage parameter in \ref{Ly's model in 2D}.



In model B, the two parameters that control the damage evolution are $\gamma_b$ and $\tau_b$. Combining these with the two nonlinear parameters in Eq.(\ref{mod-change-1d}), the parameters to be inverted are $\beta$, $\delta$, $\gamma_b$ and $\tau_b$. 

In model L, it is

\begin{equation}
  \overset{\cdot}{\alpha} = 
    \begin{cases}
      c_d \gamma_r I_2 (\xi_0  + \xi) 
      & \text{if $\xi_0  + \xi >$ 0 and $\alpha \geq$ 0}\\
      c_r \alpha \gamma_r I_2 (\xi_0  + \xi)
      & \text{if $\xi_0  + \xi \leq$ 0 and $\alpha \geq$ 0}\\
      0
      & \text{if $\alpha <$ 0}\\
    \end{cases}\,
    \label{damage-evolu-lya-ada-simp}
\end{equation}
where $C_d(\alpha) = c_d$ and $C_r(\alpha) = c_r \alpha$ in Eq. (\ref{damage-evolu-lya-ada}). This preserves the necessary components for describing the observed slow dynamics while reducing the number of parameters that need to be constrained. In this case, five parameters are related to the damage, i.e. $\xi_0$, $\gamma$, $c_d$, $c_r$, and the initial strain tensor. It is difficult to constrain all five parameters from the observations by \citet{feng2018short}. We choose the parameters following \citet{lyakhovsky1997non,lyakhovsky2016dynamic} and list them in Table \ref{inv-parameters}. We assume that $\xi_0 = 0.79$ and that the initial strain tensor in 2D is $\varepsilon_{xx0} = -e_0$, $\varepsilon_{yy0} = -e_0$ and $\varepsilon_{xy0} =\varepsilon_{yx0} = 1.46 e_0$ such that $\xi$ at the initial strain state satisfies $\xi + \xi_0 < 0$. This enables the damaged rock to heal after removing dynamic perturbations. With the above simplifications, the parameters to be inverted in model L are $\gamma_r$, $c_d$, $c_r$, and $e_0$. 


\paragraph{Bayesian inversion} \mbox{} \\
In order to quantitatively investigate the relative importance of the theoretical model parameters that explain the evolution of the observed slow dynamics, we apply Bayesian inversion to both models. We further compare the importance of first- versus  second-order nonlinearity.

The model parameters $\underset{-}{m}$ and experimental observations $\underset{-}{d}^{obs}$ are viewed as random variables in $\mathbb{M} \subset \mathbb{R}^{n_m}$ and $\mathbb{D} \subset \mathbb{R}^{n_d}$, where $n_m$ and $n_d$ are the number of model parameters and the number of observed data points. Let $G: \mathbb{R}^{n_m} \rightarrow \mathbb{R}^{n_d}$ denote the model map taking a parameter onto a model prediction.

We then aim to find the posterior, namely the conditional distribution of $\underset{-}{m}$ for a given observation $\underset{-}{d}^{obs}$. We denote the corresponding probability density function (PDF) by $\rho(\underset{-}{m}|\underset{-}{d}^{obs})$. To directly compute underlying parameters from observed data, we would have to apply the inverse of the model map $G^{-1}$. However, that inverse is not available for our models. Employing Bayes' theorem, we can reformulate the posterior in a way that, as will be detailed below, only involves the forward map $G$:

\begin{equation}
    \rho(\underset{-}{m}|\underset{-}{d}^{obs}) = \dfrac{\rho(\underset{-}{d}^{obs}|\underset{-}{m}) \rho(\underset{-}{m})}{\rho(\underset{-}{d}^{obs})} \propto \rho(\underset{-}{d}^{obs}|\underset{-}{m}) \rho(\underset{-}{m}).
    \label{bayes}
\end{equation}

We call $\rho(\underset{-}{m})$ the prior density, $\rho(\underset{-}{d}^{obs}|\underset{-}{m})$ is the likelihood that describes the probability density of measuring the observed data when $\underset{-}{m}$ is given, and $\rho(\underset{-}{d}^{obs})$ is the unconditional PDF of measuring the observed data. Broadly speaking, the posterior considers a model parameter to be likely if the parameter is plausible and its corresponding model prediction is close to observed data.

In practice, $\rho(\underset{-}{d}^{obs})$ is not available. A way to circumvent an analytical derivation is to sample the posterior with the Markov chain Monte Carlo (MCMC) method, where $\rho(\underset{-}{d}^{obs})$ cancels out due to being independent of $\underset{-}{m}$.


The prior encodes expert knowledge about what parameters might be generally plausible, not considering our specific observation. Considering the physical constraint that all model parameters are non-negative, it is $\mathbb{M}=\{\underset{-}{m} \in \mathbb{R}^{n_m} | 0 \leq m_i \leq b_i, i = 1,2,\text{\ldots},n_m \}$. $b_i$ is the assumed upper bounds of each model parameter. Another constraint on the model parameters is $\int_\mathbb{M} \rho(\underset{-}{m}) \,d\underset{-}{m} = 1$. With the above two constraints and based on the maximum entropy principle of designing the prior \citep{good1963maximum}, the data is assumed to be uniformly distributed in $\mathbb{M}$.

Since we assume measurement errors to be Gaussian, we choose the likelihood as a Gaussian distribution centered around the model prediction $G(\underset{-}{m})$:
\begin{equation}
    \rho(\underset{-}{d}^{obs}|\underset{-}{m}) = \dfrac{1}{\sqrt{(2 \pi)^n \det \underset{=}{C} } } e^{ -\dfrac{1}{2} (\underset{-}{d}^{obs}- G(\underset{-}{m}))^T \underset{=}{C}^{-1} (\underset{-}{d}^{obs}- G(\underset{-}{m})) }.
    \label{likelihood-pdf}
\end{equation}
The covariance matrix $\underset{=}{C} \in \mathbb{R}^{n_d \times n_d}$ captures the assumed variances of and the correlation between data points, encoding our knowledge about measurement accuracy. With the additional assumption that the measurements of data points are independent of each other and all have the same variance $\sigma_m^2$, we set $\underset{=}{C} = \diag{\sigma_m^2,\ldots,\sigma_m^2}$.

From Eq. (\ref{bayes}) and Eq. (\ref{likelihood-pdf}) it is clear that the posterior density can, up to an unknown constant factor, be computed point-wise. Every evaluation then requires a corresponding evaluation of the model map, i.e. one simulation run.



\paragraph{Uncertainty Quantification (UQ)} \mbox{} \\

To compute an approximation to the entire posterior distribution, we apply Markov chain Monte Carlo (MCMC). MCMC learns the posterior by randomly stepping through the parameter space. Proposed steps with high posterior probability are accepted, while low-probability proposals will likely be rejected. The distribution of model space samples obtained from the MCMC process will then converge to the desired posterior $\rho(\underset{-}{m}|\underset{-}{d}^{obs})$. The method requires a finite number of point-wise evaluations of the posterior, and can therefore numerically solve the Bayesian problem above. 

Based on the amount of information that we know about the PDEs and the cost of each forward simulation, different types of MCMC sampling algorithms may be chosen. Given that ExaHyPE does not compute the derivative of the solution with respect to the model parameters and that this is a non-trivial task for the nonlinear PDEs, we chose the AM-MCMC \citep{AMMCMC2} algorithm in this work. AM-MCMC learns an approximate variance of the posterior on the fly, automatically improving its efficiency during the run by tuning its proposals. While more complex UQ algorithms may achieve higher efficiency, AM-MCMC readily meets our accuracy and computational cost requirements.

In order to ensure that the MCMC method gives a sufficiently good approximation of the true (but unobtainable) posterior $\rho(\underset{-}{m}|\underset{-}{d}^{obs})$, we compute the Monte Carlo standard error (MCSE) as an indicator. It is defined as \citep{vehtari2021rank}

\begin{equation}
    \text{MCSE} = \sqrt{\dfrac{\text{Var}(\rho^{MC}(\underset{-}{m}|\underset{-}{d}^{obs}))}
    {S}}.
    \label{mcse}
\end{equation}
Here, $\rho^{MC}(\underset{-}{m}|\underset{-}{d}^{obs})$ is the estimated posterior, $\text{Var}(\cdot)$ is the variance of a random variable and $S$ is the number of independent samples drawn from the posterior. MCMC necessarily produces correlated samples, so the effective sample size (ESS) is applied instead of $S$. The ESS is not immediately available, but can in turn be estimated from the chain's correlated samples \citep{vehtari2021rank}.


We use the AM-MCMC implementation provided by the open-source MIT Uncertainty Quantification Library (MUQ) \citep{parno2021muq}. To couple MUQ to the forward model in the ExaHyPE simulation framework, we use the universal UQ / model interface UM-Bridge \citep{seelinger2023bridge}, which is fully supported by MUQ. For reproducibility, we provide, in the open research section, the forward model as a ready-to-run container image, that any UM-Bridge supporting UQ software can connect to. 
The MCSE estimates are provided by the ArviZ tool \citep{arviz_2019}.

\section{Results}
\label{Results}

In this section, we first show that both models capture the three phases of moduli change in DAET. Next, we analyze the dependence of the stationary damage on both the frequency and amplitude of perturbations. Finally, we quantify the sensitivity of the model parameters based on experimental data of \citet{feng2018short}.

\subsection{Theoretical models capture the observed change of modulus with strain perturbations}
\label{Theoretical models capture the change of modulus with strain during stationary damage}

\begin{figure}[ptb]
    \centering
    \includegraphics[width=1\columnwidth]{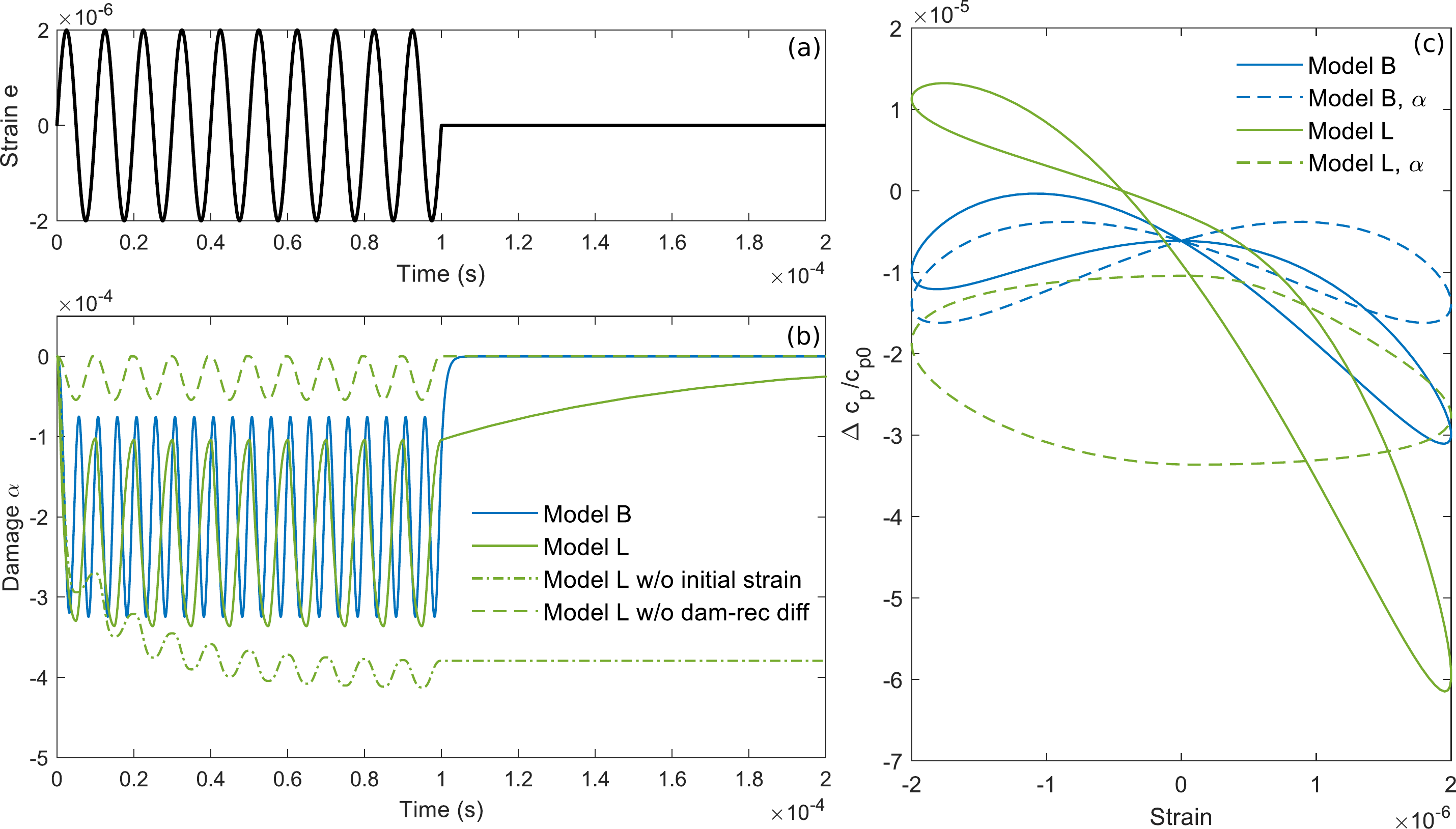}
    \caption{\small 0D evolution of the damage variables in the B and L models under dynamic perturbation that resembles a DAET experiment.
    (a) The strain perturbations added to the system. (b) Comparison of the damage conditioning and recovery of model B (blue curve), model L (solid green curve), model L without initial strain (dashed green curve), and model L without different damaging and healing evolution (dash-dotted green curve), i.e. $C_d(\alpha) = C_r(\alpha)$ in Eq. (\ref{damage-evolu-lya-ada}). Model B \citep{berjamin2017nonlinear} is described in Eq.(\ref{0d-damage}a), while model L \citep{lyakhovsky1997distributed} is described in Eq.(\ref{0d-damage}b). (c) Comparison of the change in  P-wave speed during one cycle after reaching stationary damage. $\Delta c_p / c_{p0}$ is the change of P-wave speed $\Delta c_p$ over the P-wave speed before perturbations $c_{p0}$. The P-wave speed drops of model B (the solid and the dashed blue curves) are scaled by 0.1 such that it becomes comparable to the drop of model L. The dashed blue curve shows the change of the damage variable in a strain cycle in model B, scaled by 0.1; whereas the dashed green curve shows the change of the damage variable in a strain cycle in model L, again, scaled by a factor of 0.1.}
    \label{oscillation-compare}
\end{figure}

To demonstrate the conditioning and recovery of the damage under dynamic perturbations, we here simplify both Model B and Model L to the 0D case, where damage and recovery occur under the given sinusoidal strain perturbations of a certain frequency as shown in Fig. \ref{oscillation-compare}a. As in \citet{berjamin2017nonlinear}, we assume that only one strain component $\varepsilon_{xx}$ is perturbed. The damage evolution equations for models B and L then simplify to Eqs. (\ref{0d-damage}a,b) as

\begin{table}[pt]
\centering
\caption{\small Summary of input parameters for the comparison of model B and L, with $A_0$ being the amplitude of the sinusoidal strain perturbation, $f_c$ the frequency of the strain perturbation, $\beta$ the first order nonlinearity, $\delta$ the second-order nonlinearity, $\gamma_b$ the damage energy, $\tau_b$ the evolution time scale, $\gamma_r$ the nonlinear modulus, $c_d$ the damage coefficient, $c_r$ the healing coefficient, $\xi_0$ the modulus ratio as defined in Eq. (\ref{damage-evolu-lya}), $c_r$ is the healing coefficient, $\varepsilon_{ij0}$ the different components of initial strains and $\alpha_d$ the normalization factor as defined in Eq. (\ref{0d-damage}b).}
\begin{tabular}{ M{2cm} P{2cm} P{2cm} P{1cm} P{2cm} P{2cm} P{1cm}}
\hline
 &Parameters &Values &Units &Parameters &Values &Units\\
\hline
Perturbation &$A_0$ &$2 \times 10^{-6}$ &1 &$f_c$ &$100$ &kHz \\
\hline
\multirow{2}{*}{Model B}
 &$\beta$  &$1 \times 10^{2}$ &Pa  &$\delta$ &$3 \times 10^{7}$ &Pa\\
 &$\gamma_b$ &$5 \times 10^{1}$ &Pa  &$\tau_b$ &$1 \times 10^{-6}$ &s\\
\hline
\multirow{4}{*}{Model L} 
&$c_d$ &$1.2 \times 10^{4}$ &(Pa$\cdot$s)$^{-1}$   &$\varepsilon_{xx0}$ &$-1.00 \times 10^{-6}$ &1\\
&$c_r$ &$5.0 \times 10^{6}$ &(Pa$\cdot$s)$^{-1}$   &$\varepsilon_{yy0}$ &$-1.00 \times 10^{-6}$ &1\\
&$\gamma_r$ &$8.0 \times 10^{9}$ &Pa &$\varepsilon_{xy0}$ &$1.46 \times 10^{-6}$ &1\\
&$\xi_0$  &0.79 &1 &$\alpha_d$  &$1.0 \times 10^{-4}$ &1\\
\hline
\end{tabular}
\label{0d-parameters}
\end{table}

\begin{subequations}
\begin{align}
\overset{\cdot}{\alpha} &= \dfrac{1}{\tau_b \gamma} (W - \gamma g), \\
\overset{\cdot}{\alpha} &= 
    \begin{cases}
      c_d \exp{(-\dfrac{\alpha}{\alpha_d})} \gamma_r I_2 (\xi_0  + \xi) 
      & \text{if $\xi_0  + \xi >$ 0 and $\alpha \geq$ 0}\\
      c_r \alpha \gamma_r I_2 (\xi_0  + \xi)
      & \text{if $\xi_0  + \xi \leq$ 0 and $\alpha \geq$ 0}\\
      0
      & \text{if $\alpha <$ 0}\\
    \end{cases},
\end{align}
\label{0d-damage}
\end{subequations}
where $W = (\dfrac{1}{2} - \dfrac{\beta}{3} e - \dfrac{\delta}{4} e^2) E e^2$ in Eq. (\ref{0d-damage}a), $E$ is Young's modulus, $e$ is the perturbation on $\varepsilon_{xx}$ and all other strain components are assumed to be zero. This implies that $e$ is $e_0 sin(2 \pi f_c t)$ when $t \leq 1 \times 10^{-4}$ s and becomes 0 when $t > 1 \times 10^{-4}$ s. In Eq. (\ref{0d-damage}b), $\xi$ is computed from 2D initial strain with three components $\varepsilon_{xx0}$, $\varepsilon_{yy0}$ and $\varepsilon_{xy0}$ plus the perturbation in $\varepsilon_{xx} = \varepsilon_{xx0} + e$ and the remaining strain components are assumed to be zero. 

We show the evolution of the damage variables in both models in Fig. \ref{oscillation-compare}b. We note here that the conditioning phase of Model L is subtle in Figure \ref{oscillation-compare}b but will be more pronounced under decreasing $c_d$ in Eq. (\ref{0d-damage}b). The parameters are detailed in Table \ref{0d-parameters}. We choose initial strains that satisfy $\xi(\varepsilon_{xx0},\varepsilon_{yy0},\varepsilon_{xy0}) + \xi_0 \approx -0.01 < 0$. Both Model L and Model B gradually reach the stationary damage during perturbations and recover after $e = 0$. We also show how the damage will evolve differently without adding the initial strain or without differentiating the evolution laws for damaging and healing. Without the initial strain, the stationary damage is reached at a later stage, and larger damage will be induced. In the case that $\overset{\cdot}{\alpha} = C_d \gamma_r I_2 (\xi_0  + \xi)$ irrespective of the sign of $\xi_0  + \xi$ (the green dashed curve), we observe that no damage accumulates at the end of each cycle. All the damage accumulated during the damaging phase of a cycle can always be recovered during the healing phase due to the identical evolution equations.


Fig. \ref{oscillation-compare}c shows how the change in P-wave velocity relates to the strain perturbations during the stationary phase. The change in the damage variable of models B (the dashed blue curve) and L (the dashed green curve) is of the same order of magnitude. However, the change in P-wave velocity is around one order of magnitude larger in model B than in L. One reason for this stark difference is that the modulus change is strongly influenced by the classical nonlinear stress-strain relation. Because of the different frequency contents in the evolution of $\alpha$, the relationship between the damage variable and the strain takes different shapes. Potential reasons behind such differences, which model maybe closer to reality and their physical implications will be discussed in Section \ref{Physical interpretations of the slow dynamics}.


\subsection{Amplitude- and frequency-dependent damage in 0D simulations}

Both, amplitude- and frequency-dependence of damage have been observed in the lab. Many observations show that the stationary damage grows with the magnitude and frequency of dynamic perturbations. The amplitude-dependence of damage can be resolved by various models \citep{aleshin2007microcontact,vakhnenko2005soft,FAVRIE2015221}; however, modeling the frequency dependence of damage remains challenging. The following discussion will focus  on the capabilities to also model the frequency-dependence of damage observed in laboratory experiments using models B and L.

\begin{figure}[htpb]
    \centering
    \includegraphics[width=1\columnwidth]{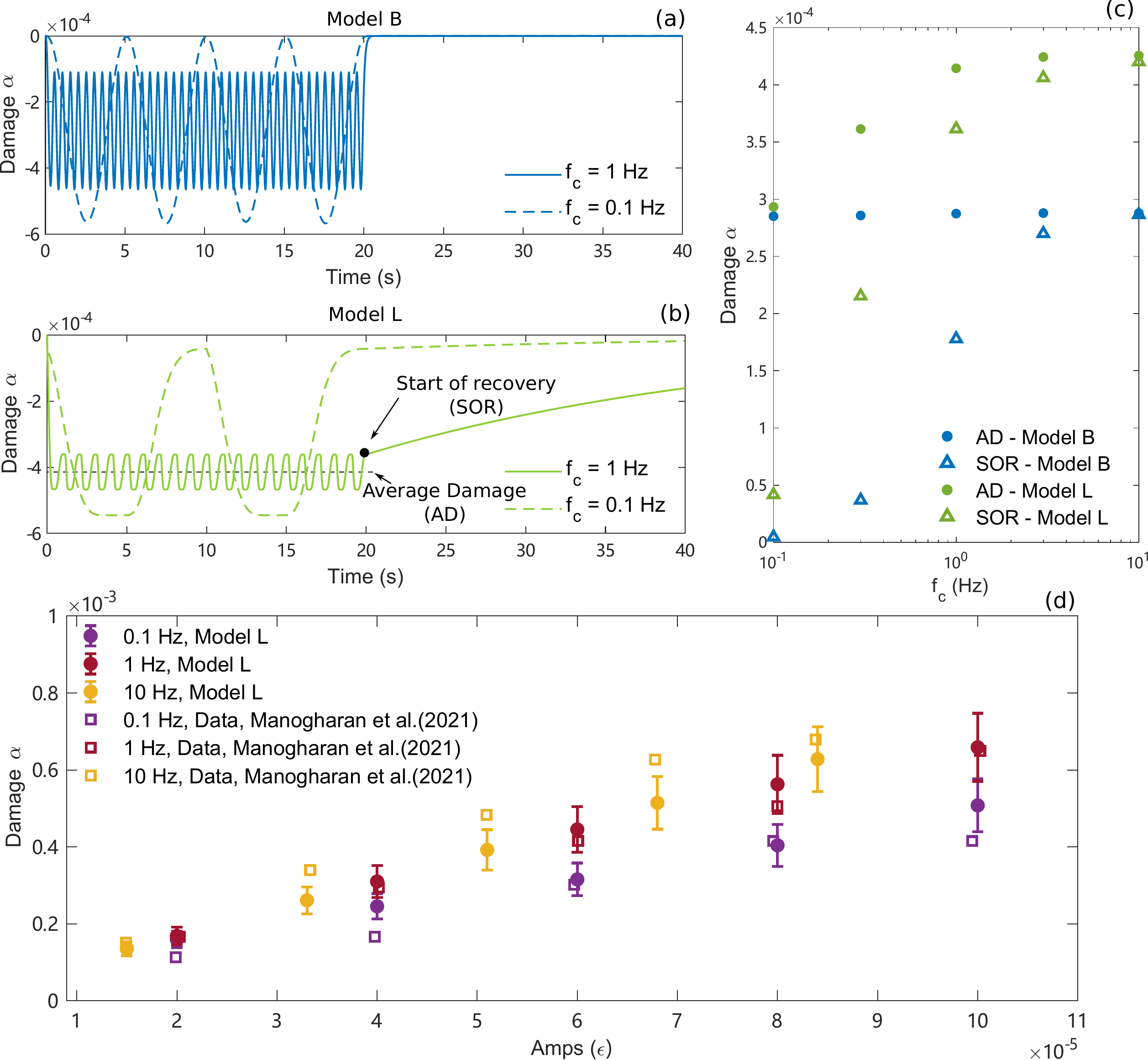}
    \caption{\small (a) Evolution of the damage variables in Model B during dynamic perturbations of a 0.1~Hz (blue dashed curve) and 1~Hz (blue solid curve) source signal. Model B \citep{berjamin2017nonlinear} is described in Eq.(\ref{0d-damage}a). (b) Evolution of the damage variables in model L during dynamic perturbations of a 0.1~Hz (green dashed curve) and 1~Hz (green solid curve) source. Model L \citep{lyakhovsky1997distributed} is described in Eq.(\ref{0d-damage}b). The start of recovery (SOR) refers to the damage value at the end of the perturbation and the average damage (AD) refers to the average value of the damage variable at the stationary phase. (c) Change of AD and SOR at the end of the perturbations under different dynamic perturbation frequencies. The blue dots indicate our results for model B while the green dots are for model L. (d) The change of AD during the stationary phase with the amplitude and the frequency of the perturbations. The rectangles show the AD measured by \mbox{\citet{manogharan2021nonlinear}} in their Fig. 8g for 5 different amplitudes and 3 frequencies (0.1 Hz in purple, 1 Hz in red and 10 Hz in yellow). The colored dots with error bars show our results from model L with the model parameter $\alpha_d$ varying between 0.4 and 0.9 in Table \mbox{\ref{afd-parameters}}. The dots represent the mean values of the model results with varying parameters; while the length of the error bars is the standard deviation of the model results. 
    }
    \label{amp-freq-damage}
\end{figure}

We show in Figs. \ref{amp-freq-damage}a-c how these dependencies evolve in our numerical simulation using the two models. In model L, both the start of recovery (SOR) and the average stationary damage (AD) increase with $f_c$, which is consistent with the observations from \citet{riviere2016frequency,manogharan2021nonlinear,manogharan2022experimental}. In model B, SOR follows a similar trend as that in model L. However, it shows a frequency independent AD. The difference between SOR and AD damage is usually not explicitly pointed out in the observations regarding the metrics for describing the damage. However, as shown in Fig. \ref{amp-freq-damage}, they can  behave very differently when quantifying the frequency-dependent damage. While AD follows different trends between models B and L, SOR consistently increases with the perturbation frequency in both models. 
We observe a frequency dependence of the variation in the damage variable.
However, the variation of the modulus appears to be frequency-independent. 
The classical nonlinearity terms $\beta$ and $\delta$ in Eq. (\ref{mod-change-1d}) and $\gamma_r$ in Eq. (\ref{damage-evolu-lya-ada-simp}) are collectively impacting on the effects of the here defined damage variable. Importantly, $\beta$, $\delta$ and $\gamma_r$ are not frequency-dependent.

A qualitative explanation of these trends in SOR can be the following. Under our assumption that the recovery rate increases with the value of the damage variable, stationarity requires a certain level of averaged damage. For higher frequencies, the amplitude of the oscillations of damage in each cycle becomes smaller. This may imply that from the same level of averaged damage, the damage value that the material reaches after one cycle (c.f., the minimum damage) will increase with frequency. 

In Fig. \ref{amp-freq-damage}d, we compare the frequency-dependent AD in model L with the measurements by \citet{manogharan2021nonlinear} using the model parameters in Table \ref{afd-parameters}. The rectangles show the AD measured by \citet{manogharan2021nonlinear} in their Fig. 8g for 5 different amplitudes and 3 frequencies (0.1 Hz in purple, 1 Hz in red, and 10 Hz in yellow). The colored dots with error bars show the results from model L with parameters shown in Table \ref{afd-parameters}. The model parameters are varied within certain ranges. The dots represent the mean values of the model results with parameter variations; while the length of error bars is the standard deviation of the model results. We find that most of the measured data points fall within the model predictions and their uncertainties are computed by varying the model parameter as in Table \ref{afd-parameters}. The average stationary damage increases almost linearly with the amplitude of perturbations \citep{johnson2005slow,manogharan2021nonlinear}. The trend of rising damage with higher perturbation frequencies is also captured by model L.

\subsection{Constraining model parameters and their uncertainties from laboratory observations}

In this section, not only the capability of the models in explaining observations is evaluated but also constraints and associated uncertainties for model parameters from laboratory experiments. 
The values of the parameters to be inverted are of vastly different magnitudes. Thus, normalization is required for the joint sensitivity analysis of all parameters.  
We assume that $\sigma_M$ in Eq. (\ref{likelihood-pdf}) is $6 \times 10^{-7}$ and $1 \times 10^{-6}$, respectively, for the inversion of Model B and Model L. 
The inversion parameters are listed in Table \ref{inv-parameters}.

\begin{table}[ht]
\centering
\caption{\small Summary of all model parameters considered in the MCMC inversion, with $\beta$ being the first-order nonlinearity, $\delta$ the second-order nonlinearity, $\gamma_b$ the damage energy, $\tau_b$ the evolution time scale, $\gamma_r$ the nonlinear modulus, $c_d$ the damage coefficient, $c_r$ the healing coefficient, $\xi_0$ the modulus ratio as defined in Eq. (\ref{damage-evolu-lya}), $c_r$ the healing coefficient and $\varepsilon_{ij0}$ the different components of initial strains. }
\begin{tabular}{ M{1cm} P{1cm} P{3cm} P{1cm} P{1cm} P{3cm} P{1cm}}
\hline
 &Param. &Values &Units &Param. &Values &Units\\
\hline
\multirow{2}{*}{Model B}
 &$\beta$  &$1.8 \times 10^{2}$ $\times$ [0,4] &Pa
 &$\delta$ &$3.0 \times 10^{8}$ $\times$ [0,4] &Pa\\
 &$\gamma_b$ &1.0 $\times 10^{ \text{[0,2]} }$   &Pa 
 &$\tau_b$ &$1.0 \times 10^{-6}$ $\times 10^{ \text{[0,3]} }$ &s\\
\hline
\multirow{4}{*}{Model L} 
&$\gamma_r$ &$4.5 \times 10^{9}$ $\times$ [0,4] &Pa  
&$\varepsilon_{xx0}$ &-$e_0$ &1\\
&$c_d$ &$1.0 \times 10^{5}$ $\times$ [0,4]      &(Pa$\cdot$s)$^{-1}$  
&$\varepsilon_{yy0}$ &-$e_0$ &1\\
&$c_r$ &$5.0 \times 10^{6}$ $\times 10^{ \text{[0,2]} }$ &(Pa$\cdot$s)$^{-1}$
&$\varepsilon_{xy0}$ &1.46$e_0$ &1\\  
&$e_0$    &$5.0 \times 10^{-8}$ $\times 10^{ \text{[0,3]} }$ &1
&$\xi_0$  &0.79 &1\\
\hline
\end{tabular}
\label{inv-parameters}
\end{table}


\begin{figure}[htpb]
    \centering
    \includegraphics[width=1\columnwidth]{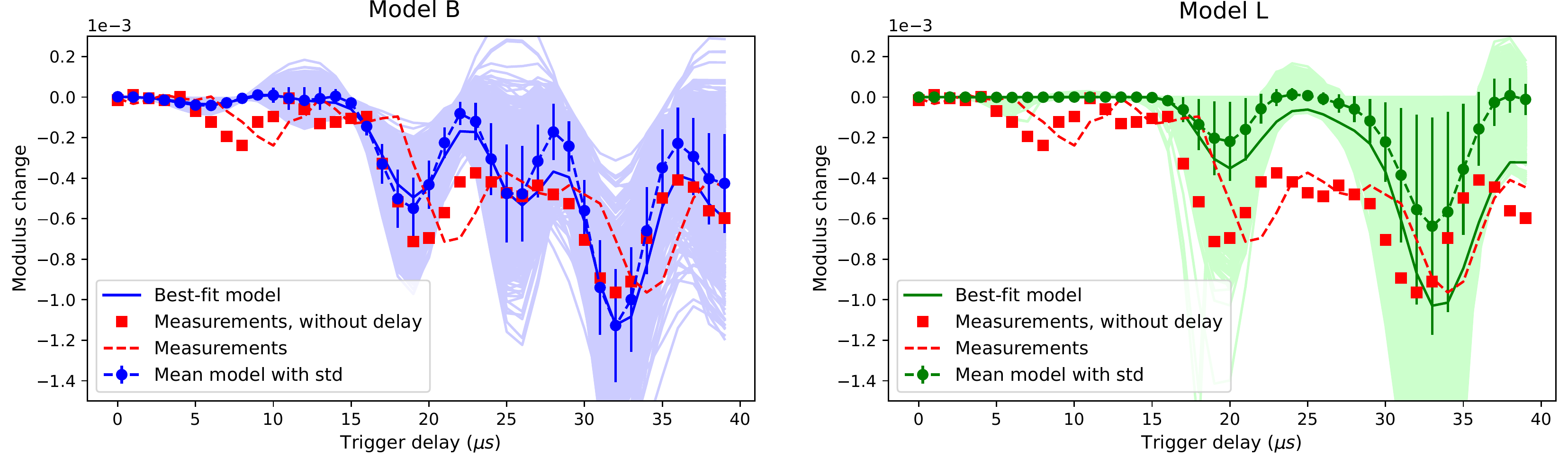}
    \caption{\small Comparison between the observations from experiments by \citet{feng2018short} and model predictions corresponding to 70,000 MCMC samples of the respective model's Bayesian posterior. The raw data that contains the delay to strain is shown with the red dashed line, while the data that removes the delay is shown with the red square dots. (a) Inversion results of model B, Eqs.(\ref{mod-change-1d}) and (\ref{0d-damage}a). The model prediction with the highest posterior probability is shown in the solid blue curve. The dashed line shows the mean prediction with error bars indicating the standard deviation. All samples' model predictions are plotted as a blue-shaded area. (b) Inversion results of model L,  Eq.(\ref{damage-evolu-lya-ada-simp}). The model prediction with the highest posterior probability is shown in the solid green curve. The dashed line shows the mean prediction with error bars indicating the standard deviation. All samples' model predictions are shown in a green shade. 
    }
    \label{model-fit}
\end{figure}

\begin{figure}[hptb]
    \centering
    \includegraphics[width=1\columnwidth]{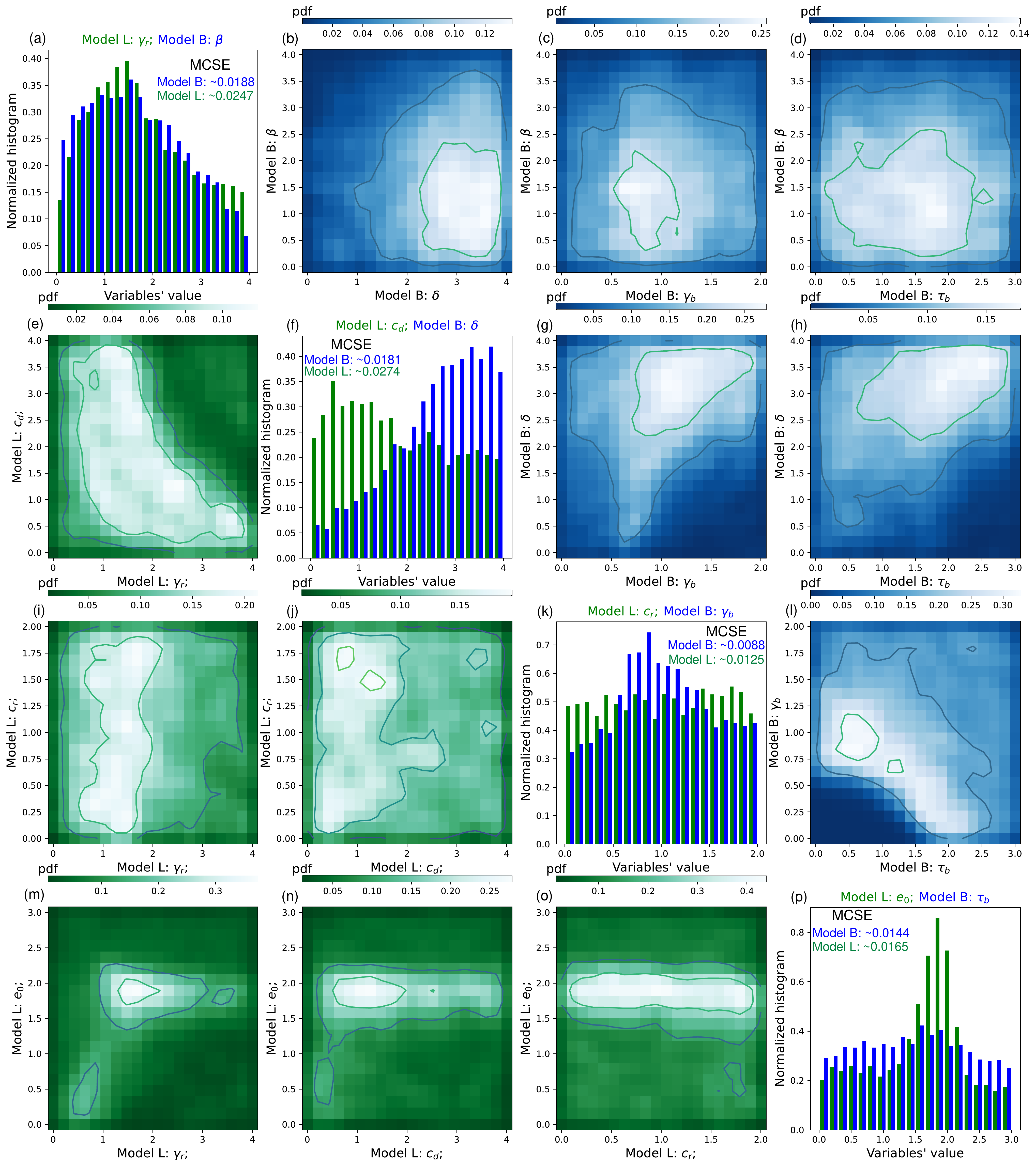}
    \caption{\small Sensitivity of different parameters in model B and in model L based on MCMC inversion. We show the one-dimension marginal probability density of the four parameters in model B (in blue) and model L (in green) on the diagonal of the plot matrix. The four parameters in model B from left to right and top to bottom are the first-order nonlinearity ($\beta$), the second-order nonlinearity ($\delta$), the damage energy ($\gamma_b$), and the evolution time scale ($\tau_b$). The four parameters in model L from left to right and top to bottom are 
    the nonlinear modulus ($\gamma$), the damage coefficient ($c_d$), the healing coefficient ($c_r$) and the initial strain level ($\varepsilon_{0}$). The two-dimension marginal probability density of all pairs of two parameters in model B is shown in the upper triangle subplots of the plot matrix. The density is shown with the color map from dark blue to white. The brighter the color, the higher the density. Similarly, the two-dimension marginal probability density of all pairs of two parameters in model L is shown in the lower triangle subplots of the plot matrix. The density is shown with the color map from dark green to white. The brighter the color, the higher the density. MCSE stands for Monte Carlo standard errors of each model parameter.}
    \label{parameter-distribution}
\end{figure}

In the MCMC runs of model B and model L, 70,000 simulations with different model parameters are sampled from the posterior. In this process, 13,826 and 10,444 proposals are accepted, respectively. The Monte Carlo standard errors (MCSE) of each parameter in models B and L are given in Figs \ref{parameter-distribution}a, f, k, and p. The inversion for each of model B and model L on a single core of the Intel i7-1165G7 processor takes around 10 hours.

We compare the simulation results and the observations from the experiment in Fig. \ref{model-fit}. Simulation results from the best-fit parameter set of Model B and Model L match the observations with the correlation coefficients of 0.91 (Model B) and 0.83 (Model L), respectively. 
While model L does not fit the data as well as model B, it reproduces the delay of the peak in modulus change compared to the peak in strain. This is not explained by either model B or the visco-elastic formulation proposed in \citet{feng2018short}. If only the nonlinear stress-strain relationship is considered, the largest modulus drop occurs when the tensile strain peaks. In model L, when the tensile strain reaches the peak, damage development will not cease. It only stops upon the strain becoming sufficiently compressed. This further damage that lasts for around 1/4 of the cycle leads to a larger modulus drop after peak tensile stress as is observed in the laboratory data.


The marginal probability distributions of the laboratory-constrained model parameters of models B and L are shown in Fig. \ref{parameter-distribution}. The histograms in blue on the diagonal of Fig. \ref{parameter-distribution} show the one-dimensional marginal distribution of the model parameters in model B. The values with the highest marginal probability for the normalized $\beta$, $\delta$, $\gamma_b$ and $tau_1$ are 1.10, 3.21, 0.88, and 1.75. The two-dimensional marginal distribution of each pair of two parameters is shown on the upper triangle of Fig. \ref{parameter-distribution}. The first-order nonlinearity parameter $\beta$ remains relatively independent of other parameters. In distinction, the correlations among the other three parameters are significantly stronger. We also infer that with higher damage energy ($\gamma_b$), the second-order nonlinearity will more likely be larger whereas the time scale for damage evolution $\tau_b$ decreases.

Similarly, the histograms in green on the diagonal of Fig. \ref{parameter-distribution} show the one-dimension marginal distribution of the model parameters in model B. The values with the highest marginal probability for the normalized $\gamma_r$, $c_d$, and $e_0$ are 1.30, 0.50, and 1.80. We find that the healing coefficient $c_r$ is not well constrained by the laboratory observations. The two-dimensional marginal distribution of each pair of two parameters is shown on the lower triangle of Fig. \ref{parameter-distribution}. We observe a negative correlation between the nonlinear modulus $\gamma_r$ and the damage coefficient $c_d$. We relate this effect to the lack of resolution in constraining the healing coefficient. According to Figs. \ref{parameter-distribution}e and j, the trend of damage increase due to larger damage coefficient $c_d$ is better compensated for by the decrease of the nonlinear modulus $\gamma_r$ than by the increase of the healing coefficient $c_r$.





\section{Discussion}
\label{Discussion}


In this paper, we analyse the behavior of two physical models to simulate nonlinear elastic wave propagation and compare them with laboratory observations. Several interesting aspects are worth further discussing here: (1) We find that Model L can reproduce the increase of average damage (AD) with the frequency of strain perturbations during the steady state of damage. We will here revisit this main advantage and interpret it from both mathematical and physical points of view. (2) We will also discuss how the influence of the damage variable on the wave speeds can be similar to second-order nonlinearity. (3) In model L, one of the key assumptions required for recovery after removing perturbations is the existence of initial strain. We will discuss the validity of this assumption.

\subsection{The frequency-dependence of damage}

Recent studies find a general trend of increasing average damage with the excitation frequency during the stationary phase \citep{riviere2016frequency,manogharan2022experimental,manogharan2021nonlinear}. However, the underlying mechanism is not well understood. While \citet{riviere2016frequency} favors models with rate/time dependencies, not every such model also results in frequency-dependent stationary damage. In model B, the average damage is $ \dfrac{E}{4 \gamma_b} A^2 + \mathcal{O}(A^4)$ \citep{berjamin2017nonlinear}, where $A$ is the strain amplitude of perturbations and is not related to the frequency (see also Fig. \ref{amp-freq-damage}). In model L, frequency-dependent average damage is observed and follows the same trend as  the measurements (c.f., the green dots in Fig. \ref{amp-freq-damage}c). 

We here interpret these findings analytically (see \ref{Frequency dependence of Ly's model}). We show that in model L, the decrease of the damage coefficient with damage and the increase of the healing coefficient contribute differently to the frequency-dependent average stationary damage.  We can also show that damage at the stable state can increase with the frequency of the dynamic perturbation. Specifically, with increasing healing coefficient ($C_r(\alpha) = c_r \alpha$), the average damage tends to decrease with frequency, i.e. Eq. (\ref{ad-increaseheal}); while with decreasing damage coefficient ($C_d(\alpha) = c_d \exp{(\alpha/\alpha_d)}$), the average stationary damage becomes larger as the frequency increases, i.e. Eq. (\ref{ad-decreasedamage}). 


\subsection{Damage vs. second-order nonlinearity contributions to co-seismic changes in wave speed}


Forming a "bow-tie" loop relation between the wave speed variation and the strain perturbation requires that the time series of the modulus change contain a frequency component that doubles the frequency of strain oscillation. In model B, there are two sources of such double frequency. The first mechanism is second-order nonlinearity, as has been recognized in previous analysis \citep{sens2019model}. The second mechanism contributing to the double frequency is related to damage and healing within a cycle. The rock is damaged when the absolute value of the strain increases while it heals otherwise. This leads to two cycles of damage and healing inside one cycle of strain perturbation. 

In model L, neither the second-order nonlinearity ($\delta$) nor the double damage-healing cycles exist. The rock damages when the tensile strain is large enough that $\xi_0+\xi > 0$ and it heals otherwise. This implies that the oscillation of the damage variable has the same frequency as the strain perturbations. This manifests in the ellipse-shaped (the green-dotted curve) dependence of the damage variable on strain in Fig. \ref{oscillation-compare}c. Instead, the mechanism in model L that adds the double frequency to the damage evolution is the change of wave speed with the amplitude of the strain, coming from the third non-quadratic term in Eq. (\ref{inter-energy-lya}). 

We caution that damage effects may be confused with the effects of second-order nonlinearity in observations of "bow-tie" loops. In the framework of this paper, we introduce a damage variable to explain the conditioning and the recovery of the wave speed. However, in addition to slow dynamics, the damage variable can also influence the measured wave speed in a similar way as classical nonlinearity. Such confusion may also appear in comparisons of modulus variation as in Fig. \ref{parameter-distribution}. The third parameter (the damage energy $\gamma_b$) in model B determines the magnitude of damage: the smaller the damage energy, the larger the magnitude of the damage. We conclude that the positive correlation between $\gamma_b$ and the second-order nonlinearity ($\delta$) implies that the effect of the damage evolution on the modulus is comparable to that of the second-order nonlinearity. 
However, it is possible to discern the fundamental difference in the effects of the damage variable versus second-order (classical) nonlinearity. The latter will lead to the same strain-dependent modulus change, irrespective of the loading rate. However, the change of modulus due to the damage variable will always be rate-dependent, as given in Eq. (\ref{damage-evolu}). 

\subsection{Physical interpretation of both damage models}
\label{Physical interpretations of the slow dynamics}

Whether damage evolution discriminates between compression and extension can lead to different interpretation of the physical mechanisms underlying the evolution of the damage variables. Different hypotheses have been proposed relating damage to friction \citep{aleshin2007microcontact} or to adhesion \citep{lebedev2014unified}. 
Since the sign of $\gamma_b g$ in Eq. (\ref{inter-energy-berjamin}) is not related to compression or extension of strain, the healing and damage terms of model B are clearly separated. The healing term becomes larger than zero once damage occurs irrespective of the strain state. The healing term is simultaneously increasing with the accumulation of damage. Once the healing term becomes large enough, damage accumulation reaches a quasi-static state. 

In contrast, the damage evolution of the material in model L differentiates between compressive and extensive deformation. Here, rock only heals when the material is sufficiently compressed ($\xi + \xi_0 < 0$). The quasi-static state is reached due to increasing healing speed. 
\citet{lebedev2014unified} proposed a possible explanation of slow dynamics being related to the thermal processes due to adhesion at the contacts of rock grains. The adhesion potential \citep{jacob1992intermolecular} is non-symmetric for compression and extension. The mechanism of adhesion therefore also favors a model that considers different effects of damage and healing under compression and extension.


While both models explain the slow dynamics mathematically, the physical interpretation of the L model has richer physical implications. First, each term in the internal energy of the L model has a physical meaning, and the formulation strictly follows the laws of thermodynamics. The healing is related to $\gamma I_1 \sqrt{I_2}$, a term that comes from the opening and closing of micro-cracks \citep{lyakhovsky1997distributed}. At the microscopic scale, this may be interpreted as a re-attachment of asperities at contact surfaces. Second, as shown in Fig. \ref{parameter-distribution}, the initial strain is relatively well resolved to be around $3 \times 10^{-6}$. The existence of this non-negligible initial strain may be related to the cohesive contact or to the thermal deformation of rocks. The typical thermal expansion coefficient of rock is of the order of $10^{-5} K^{-1}$ \citep{kirk2012structure}, where $K$ is the unit of the absolute temperature. This means a change of less than 1 $K$ may equate to the here-constrained magnitude of the initial strain. Although more observations are required to confirm this interpretation, the existence of initial strain possibly explains why slow dynamics typically become prominent when the dynamic strain is larger than $10^{-6}$ but not for smaller values. We also  note that the stationary point is reached due to the increasing speed of healing under compression  with higher accumulated damage. This increase is not necessarily linear. It is here only assumed to be linear for simplicity and demonstration purposes. A physically motivated expression requires specific experiments to constrain this behavior.



\subsection{Limitations}



The damage variable capsulizes changes in material stress-strain relations due to the physical processes, such as adhesion \citep{lebedev2014unified}, plastic deformation at grain contacts \citep{lieou2017slow} or friction \citep{aleshin2007microcontact}, at meso- and microscopic scales. Compared to the explanation of \citet{lebedev2014unified}, the accumulation of damage during the extension of material might be related to the detachment at the contacts of asperities; whereas the recovery (decrease of the damage variable value) might be associated with re-attachment and with asperities that gradually shift from the secondary stationary state to the main stationary state. However, developing a stringent mathematical framework connecting the damage variable at the macroscopic scale with microscopic physical processes is challenging and currently elusive.


Model L does not match the modulus change equally well as Model B, especially during rock compression. In Model L, the increase in wave speed due to the third term $\gamma I_1 \sqrt{I_2}$ during compression cancels out the wave-speed drop from damage. Future modification of the L Model may mitigate the influence of its third term on the change in wave speed while preserving differentiating between tensile and compressive strains.

In this work, we analyze two damage models from 0D to 2D. To compare our 2D simulation results with laboratory observations, we simplify some of the model parameters acknowledging the limited amount of available observational data. One aspect that will require further investigation for modeling slow dynamics in 2D and 3D is the full response of material damage from strain perturbations including different components of the strain/stress tensor. \citet{lott2017nonlinear} proposed a formulation that connects a scalar damage variable to the stiffness tensor. The resulting changes in P- and S-wave speed in response to different modes of perturbations qualitatively match observations. In future work, this relationship may be combined into the internal energy formulation in Eq.(\ref{gibbs}) to potentially refine our dynamic understanding of how stationary damage is reached.


The achieved quantitative match of our modeling results and laboratory observations demonstrate the potential of both proposed models to capture natural co-seismic damage of rocks. As a next step, the nonlinear models can be implemented in  large-scale 3D wave solvers, e.g., SeisSol
(\mbox{\href{https://seissol.org}{https://seissol.org}}) or ExaHyPE \mbox{\citep{reinarz2020exahype}}. The associated major challenges stem from the nature of nonlinear hyperbolic partial differential equations. Solutions can become discontinuous during the propagation of waves even when the initial conditions are smooth \citep{leveque2002finite}. Numerical methods may try to resolve such dynamic discontinuities by introducing numerical diffusion. This comes with different criteria for numerical stability and typically requires higher spatial and temporal resolution. It leads to computationally more expensive schemes. However, to simulate co-seismic damage and post-seismic recovery as observed in the field, 3D simulations will be indispensable to help bridge the gap between the laboratory and the field scales.

\section{Conclusions}
\label{Conclusions and outlook}

We demonstrate the applicability and compare two models that explain the observed non-classical nonlinear behaviors of rocks, an internal variable model (IVM) and a continuum damage model (CDM).
The analyzed IVM, model B, is proposed by \citet{berjamin2017nonlinear}, while the CDM, model L, is adapted from \citet{lyakhovsky1997distributed}. Using both physical models, we numerically simulate nonlinear wave propagation in rocks with fast and slow dynamics with the discontinuous Galerkin (DG) method in 0D, 1D, and 2D.

We compare the simulation results with two sets of experiments. In co-propagating acousto-elastic testing, the change of modulus in laboratory observation has a higher correlation coefficient with simulation results using the model by \citet{berjamin2017nonlinear} than those of the model that we adapted from \citet{lyakhovsky1997distributed}. However, only the latter model explains both, the observed delay of modulus variation relative to strain and the frequency-dependent damage.

We quantify the possibilities to constrain nonlinear model parameters from co-propagating acousto-elastic testing observations using Adaptive Metropolis Markov chain Monte Carlo (AM-MCMC). From the joint posterior distribution of the model's parameter space, we demonstrate that nonlinear parameters can be resolved but that the associated uncertainties vary. We find that the effects on wave-speed changes from the second-order nonlinearity and from the damage variable can be very similar.
The evolution time scale ($\tau_b$) in the model of \citet{berjamin2017nonlinear} and the $healing$ coefficients in the adapted model from \citet{lyakhovsky1997distributed} are particularly challenging to resolve.

We conclude that the quantitative match between both models and the laboratory observations justify the applicability of these models in describing the phenomena of slow dynamics. 
We show that in model L, the slow dynamics are prominent only if the strain perturbation is large enough compared to the initial strain level in the material. We discuss that such conditions may be related to a "magical" strain level of around $10^{-6}$, below which the slow dynamics are not observed.
%
Future nonlinear damage modeling using either physical model in 3D highly-scalable software for seismic wave propagation simulations will allow comparison to field-scale observations and account for natural complexities such as complex surface topography and subsurface heterogeneities.


\acknowledgments
The authors are grateful for helpful and inspiring discussions with Christoph Sens-Schönfelder, Harold Berjamin, Vladimir Lyakhovsky, Dave A. May, Anne Reinarz, Michael Dumbser, Evgeniy Romenskiy, Yehuda Ben-Zion, Sebastian Wolf and Parisa Shokouhi.

This project has received funding from the European Union's Horizon 2020 research and innovation programme under the Marie Skłodowska-Curie grant agreement No 955515 – SPIN ITN (www.spin-itn.eu). AAG acknowledges additional support from European Union’s Horizon 2020 Research and Innovation Programme (TEAR grant No. 852992) and Horizon Europe (ChEESE-2P grant No. 101093038, DT-GEO grant No. 101058129 and Geo-INQUIRE grant No.101058518), the National Science Foundation (grant No. EAR-2121666), the National Aeronautics and Space Administration (80NSSC20K0495) and the Southern California Earthquake Center (SCEC award 22135).

\section*{Open research}

The combination of the AM-MCMC algorithm of MUQ \citep{parno2021muq} and the forward modeling in ExaHyPE \citep{reinarz2020exahype} is implemented with UM-Bridge \citep{seelinger2023bridge}. The detailed description of these packages, the algorithms therein, as well as the code of our implementation are provided in the following repository: 
\\ \href{https://zenodo.org/badge/latestdoi/551506661}{https://zenodo.org/badge/latestdoi/551506661}.


\appendix
\section{Model L in 2D}
\label{Ly's model in 2D}

The model of \citet{lyakhovsky1997distributed}, with plane strain assumption, can be written in 2D as a set of hyperbolic PDEs as

\begin{equation}
    \dfrac{ \partial \underset{-}{q} }{ \partial t } =
    \dfrac{ \partial \underset{-}{F} }{ \partial x } +
    \dfrac{ \partial \underset{-}{G} }{ \partial y } +
    \underset{-}{s},
    \label{pde-lya}
\end{equation}
where 
\begin{align*}
    \underset{-}{q} &= (\varepsilon_{xx},\varepsilon_{yy},\varepsilon_{xy},v_x,v_y,\alpha)^T,\\
    \underset{-}{F} &= (v_x,0,\dfrac{1}{2}v_y,\sigma_{xx}/\rho,\sigma_{xy}/\rho,0)^T,\\
    \underset{-}{G} &= (0,v_y,\dfrac{1}{2}v_x,\sigma_{xy}/\rho,\sigma_{yy}/\rho,0)^T,\\
    \underset{-}{s} &= (0,0,0,0,0,\overset{\cdot}{\alpha}),
\end{align*}
and $\overset{\cdot}{\alpha}$ is given in Eq. (\ref{damage-evolu-lya-ada}). The stress-strain relationship is $\sigma_{ij} = (\lambda I_1 - \gamma \sqrt{I_2}) \delta_{ij} + (2 \mu - \gamma \dfrac{I_1}{\sqrt{I_2}}) \varepsilon_{ij}$. It is therefore derived that

\begin{equation}
    \dfrac{ \partial \underset{-}{F} }{ \partial \underset{-}{q} } = 
    \begin{bmatrix} 
0 &0    &0    &0    &1    &0    &0\\ 
0 &0    &0    &0    &0    &0    &0\\
0 &0    &0    &0    &0    &1/2    &0\\
Q_{11} &Q_{12}    &Q_{13}    &0    &0    &0    &D_1\\
Q_{21} &Q_{22}    &Q_{23}    &0    &0    &0    &D_2\\
0 &0    &0    &0    &0    &0    &0
\end{bmatrix},
    \label{jaco-lya}
\end{equation}
where
\begin{align*}
    \rho Q_{11} &= (\lambda + 2\mu) - \gamma (2 \dfrac{\varepsilon_{xx}}{\sqrt{I_2}} +
    \dfrac{I_1 (\varepsilon_{yy}^2 + 2\varepsilon_{xy}^2) }{I_2 \sqrt{I_2}} ),\\
    \rho Q_{12} &= \lambda - \gamma (\dfrac{I_1}{\sqrt{I_2}} - 
    \dfrac{I_1 \varepsilon_{xx} \varepsilon_{yy} }{I_2 \sqrt{I_2}} ),\\
    \rho Q_{13} &= - \gamma (\dfrac{2 \varepsilon_{xy}}{\sqrt{I_2}} - 
    \dfrac{2 I_1 \varepsilon_{xx} \varepsilon_{xy} }{I_2 \sqrt{I_2}} ),\\
    \rho Q_{21} &= - \gamma (\dfrac{\varepsilon_{xy}}{\sqrt{I_2}} - 
    \dfrac{I_1 \varepsilon_{xx} \varepsilon_{xy} }{I_2 \sqrt{I_2}} ),\\
    \rho Q_{22} &= - \gamma (\dfrac{\varepsilon_{xy}}{\sqrt{I_2}} - 
    \dfrac{I_1 \varepsilon_{yy} \varepsilon_{xy} }{I_2 \sqrt{I_2}} ),\\
    \rho Q_{23} &= 2 \mu - \gamma I_1 \dfrac{\varepsilon_{xx}^2 + \varepsilon_{yy}^2}{I_2 \sqrt{I_2}} .
\end{align*}

With the same method as in \citet{berjamin2019plane}, the P-wave velocity in the x direction reads

\begin{equation}
    c_p = \dfrac{1}{2} \sqrt{ 2 Q_{11} + Q_{23} + \sqrt{ (2 Q_{11} - Q_{23})^2 +8 Q_{13} Q_{21} } },
    \label{cp-lya}
\end{equation}

\section{Analytical interpretation of the frequency dependence of model L}
\label{Frequency dependence of Ly's model}

Here we derive analytical solutions to the frequency-dependent damage of the model of \citet{lyakhovsky1997distributed}. We make the following simplifying assumptions: First, we assume that only $\varepsilon_{xx}$ is perturbed by $e = e_0 sin(\omega_c t)$, as in Fig. \ref{oscillation-compare}. As in Table \ref{0d-parameters}, $\xi(\underset{=}{\varepsilon_0}) + \xi \approx 0.01$ and is assumed to be zero at the initial strain level. Further, we approximate $I_2 (\xi_0 + \xi) = R_0 e + \mathcal{O}(e^2) \approx R_0 e$. 

\begin{itemize}
    \item \textbf{Increasing healing coefficient with damage}
\end{itemize}

In the case when the healing coefficient linearly increases with the damage variable, Eq. (\ref{damage-evolu-lya-ada}) is simplified as

\begin{equation}
  \overset{\cdot}{\alpha} = 
    \begin{cases}
      C_d \gamma_r R_0 e = \dfrac{c_1}{e_0} e
      & \text{if $e >$ 0 and $\alpha \geq$ 0}\\
      C_r \alpha \gamma_r  R_0 e = \dfrac{k_1}{e_0} \alpha e
      & \text{if $e \leq$ 0 and $\alpha \geq$ 0}\\
      0
      & \text{if $\alpha <$ 0}\\
    \end{cases}.
    \label{damage-evolu-lya-ada-app1}
\end{equation}

Consider one cycle of perturbation where $t \in [0,2\pi/\omega_c]$ and the damage variable at the beginning of the cycle is $\alpha_0$. We then derive that the maximum damage at $t=\pi/\omega_c$ is $\alpha_{max} = \alpha_0 + 2c_1/\omega_c$ at the end of the cycle, $\alpha_t = \alpha_{max} \exp{(-2k_1/\omega_c)}$. At the dynamically stable state, $\alpha_t = \alpha_0$. The average of the damage variable in a cycle is approximated by 

\begin{align}
    <\alpha> \approx 0.5 ( \alpha_{0} + \alpha_{max}) =\dfrac{c_1}{\omega_c} \dfrac{1+e^{-2k_1/\omega_c}}{1-e^{-2k_1/\omega_c}},
    \label{ad-increaseheal}
\end{align}
which is monotonously decreasing with frequency.


\begin{itemize}
    \item \textbf{Decreasing damaging coefficient with damage}
\end{itemize}

In the case when the damage coefficient exponentially decreases with the damage variable, Eq.(\ref{damage-evolu-lya-ada}) is simplified as

\begin{equation}
  \overset{\cdot}{\alpha} = 
    \begin{cases}
      C_d \gamma_r R_0 \exp{(-\alpha/\alpha_d)} e = \dfrac{c_1}{e_0} \exp{(-\alpha/\alpha_d)} e
      & \text{if $e >$ 0 and $\alpha \geq$ 0}\\
      C_r \gamma_r  R_0 e = \dfrac{c_2}{e_0} e
      & \text{if $e \leq$ 0 and $\alpha \geq$ 0}\\
      0
      & \text{if $\alpha <$ 0}\\
    \end{cases}.
    \label{damage-evolu-lya-ada-app2}
\end{equation}

Following a similar consideration as above, we derive

\begin{align}
    \exp{(\alpha_0/\alpha_d)} &= \dfrac{K}{1-K} \dfrac{2 c_1}{ \omega_c \alpha_d }, 
    \label{fdd-eq1}\\
    \exp{(\alpha_{max}/\alpha_d)} &= \exp{(\alpha_0/\alpha_d)} + \dfrac{2 c_1}{\alpha_d \omega_c},
    \label{fdd-eq2}
\end{align}
where $K = \exp{(-\dfrac{2 c_2}{\alpha_d \omega_c})}$. Multiplying Eq. (\ref{fdd-eq1}) and (\ref{fdd-eq2}), it is derived that
\begin{align}
    \exp{( \dfrac{\alpha_0+\alpha_{max}}{\alpha_d})} &= \dfrac{K}{(1-K)^2} (\dfrac{2 c_1}{ \omega_c \alpha_d })^2.
    \label{ad-decreasedamage}
\end{align}
with the right-hand side monotonously increasing with frequency.

Analytical analysis of the combination of the two cases in Eq. (\ref{damage-evolu-lya-ada-app1}) and (\ref{damage-evolu-lya-ada-app2}) is challenging. But within the parameter space used in Fig. \ref{amp-freq-damage}, we can show that the damage at the stable state can increase with the frequency of the dynamic perturbation.

\section{Model parameters used to compare with the observed amplitude-frequency dependence}
\label{Model parameters used to compare with the observed amplitude-frequency dependence}

The evolution of damage follows Eq. (\ref{damage-evolu-lya-ada}) with $C_d(\alpha) = c_d \exp{(-\dfrac{\alpha}{\alpha_d \sqrt{I_2} })}$ and $C_r(\alpha) = c_r \alpha$. We adopt a slight change in $C_d(\alpha)$ for a better fit of the data. Our adapted damage evolution is still subject to the restrictions from the laws of thermodynamics in Eq. (\ref{disspate-alpha}), i.e., ensuring that the entropy is increasing in a closed system.

\begin{table}[hpt]
\centering
\caption{\small Summary of parameters for the comparison of model B and L, with $A_0$ being the amplitude of the sinusoidal strain perturbation, $f_c$ the frequency of the strain perturbation, $\beta$ the first order nonlinearity, $\delta$ the second-order nonlinearity, $\gamma_b$ the damage energy, $\tau_b$ the evolution time scale, $\gamma_r$ the nonlinear modulus, $c_d$ the damage coefficient, $c_r$ the healing coefficient, $\xi_0$ the modulus ratio as defined in Eq. (\ref{damage-evolu-lya}), $c_r$ the healing coefficient, $\varepsilon_{ij0}$ the different components of initial strains and $\alpha_d$ the normalization factor as defined in \ref{Model parameters used to compare with the observed amplitude-frequency dependence}.}
\begin{tabular}{ M{2cm} P{1cm} P{2cm} P{1cm} P{1cm} P{2cm} P{1cm}}
\hline
 &Para. &Values &Units &Para. &Values &Units\\
\hline
Perturbation &$A_0$ &$6 \times 10^{-5}$ &1 &$f_c$ &0.1 or 1 &Hz \\
\hline
\multirow{2}{*}{Model B}
 &$\beta$  &$1 \times 10^{2}$ &Pa  &$\delta$ &$3 \times 10^{7}$ &Pa\\
 &$\gamma_b$ &$3 \times 10^{4}$ &Pa  &$\tau_b$ &$1 \times 10^{-1}$ &s\\
\hline
\multirow{4}{*}{Model L} 
&$c_d$ &$5.0 \times 10^{0}$ &(Pa$\cdot$s)$^{-1}$   &$\varepsilon_{xx0}$ &$-2.00 \times 10^{-6}$ &1\\
&$c_r$ &$5.0 \times 10^{-2}$ &(Pa$\cdot$s)$^{-1}$   &$\varepsilon_{yy0}$ &$-2.00 \times 10^{-6}$ &1\\
&$\gamma_r$ &$8.0 \times 10^{9}$ &Pa &$\varepsilon_{xy0}$ &$6.78 \times 10^{-6}$ &1\\
&$\xi_0$  &0.39 &1 &$\alpha_d$  &$[0.4,0.9]$ &1\\
\hline
\end{tabular}
\label{afd-parameters}
\end{table}

\bibliography{references}

%
%
%
%
%

\end{document}